\documentclass[a4paper,12pt,notitlepage]{article}
\usepackage{amsfonts,amssymb,amsmath,amsthm,mathtools,hyperref,colonequals}
\usepackage{caption}
\usepackage{nicefrac}
\usepackage{cite}
\usepackage[utf8]{inputenc}
\usepackage{tabularx}
\usepackage{wasysym}
\usepackage{bbm}
\usepackage{color}
\usepackage{framed}
\definecolor{shadecolor}{rgb}{1,0.9,0.7}
\usepackage{graphicx}
\usepackage{shuffle}
\usepackage{url}
\usepackage{verbatim}
\usepackage{mathrsfs}

\numberwithin{equation}{section}

\theoremstyle{definition}
\newcommand{\beqa}{\begin{eqnarray}}
\newcommand{\eeqa}{\end{eqnarray}}
\newcommand{\beq}{\begin{equation}}
\newcommand{\eeq}{\end{equation}}

\newcommand{\fP}{\mathsf{P}}
\newcommand{\fp}{\mathsf{p}}
\newcommand{\ftau}{\mathsf{t}}
\newcommand{\fTau}{\mathsf{T}}
\newcommand{\fq}{\mathsf{q}}

\newcommand{\calF}{\mathscr{F}}
\newcommand{\calS}{\mathcal{S}}

\newcommand{\calA}{\mathcal{A}}

\newcommand{\calO}{\mathcal{O}}
\newcommand{\calZ}{\mathcal{Z}}
\newcommand{\calN}{\mathcal{N}}

\DeclareMathOperator{\Tr}{Tr}

\newcommand{\melement}[3]{\ensuremath{\left< #1\left| \, #2\,\right| #3\right>}}

\newcommand{\abra}[2]{\ensuremath{\langle #1 #2\rangle}}
\newcommand{\sbra}[2]{\ensuremath{[ #1 #2]}}

\newcommand{\sabra}[2]{\ensuremath{\langle #1 #2\rangle}} 

\newcommand{\dd}{\mathrm{d}}
\newcommand{\DD}{\mathrm{D}}
\newcommand{\AAA}{\mathcal{A}}
\newcommand{\WWW}{\mathcal{W}}

\newcommand{\PPP}{\mathbf{F}}
\newcommand{\la}{\lambda}

\newcommand{\m}{\textbf{m}}\newcommand{\bm}{\bar{\textbf{m}}}
\newcommand{\n}{\textbf{n}}

\newcommand{\xdot}{\boldsymbol{\cdot}}

\newcommand{\notA}{B}

\newcommand{\e}{\operatorname{e}}
\newcommand{\ri}{i}
\newcommand{\rj}{j}

\newcommand{\eqndot}{\, .}
\newcommand{\eqncom}{\, ,}
\newcommand{\eqnsem}{\, ;}
\newcommand{\YM}{{\mathrm{\scriptscriptstyle YM}}}
\DeclareMathOperator{\phaneq}{\phantom{{}=}}

\makeatletter
\DeclareRobustCommand*{\bfseries}{%
  \not@math@alphabet\bfseries\mathbf
  \fontseries\bfdefault\selectfont
  \boldmath
}

\makeatother

\begin{document}

\thispagestyle{empty}
\setcounter{page}{0}
\begin{flushright}\footnotesize
\texttt{HU-Mathematik-2016-06}\\
\texttt{HU-EP-16/10}\\
\texttt{MITP/16-027}\\
\vspace{0.5cm}
\end{flushright}
\setcounter{footnote}{0}

\begin{center}
{\huge{
\textbf{All Tree-Level MHV Form Factors \vspace{2mm} \\ 
 \vspace{2mm} 
in $\calN=4$ SYM from Twistor Space
}
}}
\vspace{10mm}

{\sc 
Laura Koster$^{a}$, Vladimir Mitev$^{a,b}$,\\ Matthias Staudacher$^a$, Matthias Wilhelm$^{a,c}$}\\[5mm]

{\it $^a$Institut f\"ur Mathematik, Institut f\"ur Physik und IRIS Adlershof,\\ Humboldt-Universit\"at zu Berlin,\\
Zum Gro{\ss}en Windkanal 6,  12489 Berlin, Germany
}\\[2.5mm]
{\it $^b$PRISMA Cluster of Excellence, Institut f\"ur Physik, WA THEP,\\ 
Johannes Gutenberg-Universit\"at Mainz,\\
Staudingerweg 7, 55128 Mainz, Germany
}\\[2.5mm]
{\it $^c$Niels Bohr Institute, Copenhagen University,\\
Blegdamsvej 17, 2100 Copenhagen \O{}, Denmark
}\\[5mm]

\{\texttt{laurakoster@physik.hu-berlin.de},
\texttt{vmitev@uni-mainz.de},
\texttt{matthias@math.hu-berlin.de},
\texttt{matthias.wilhelm@nbi.ku.dk}\}\\
[15mm]

\textbf{Abstract}\\[2mm]
\end{center}
We incorporate all gauge-invariant local composite operators into the twistor-space formulation of $\mathcal{N}=4$ SYM theory, detailing and expanding on ideas we presented recently in \cite{Koster:2016ebi}.
The vertices for these operators contain infinitely many terms and we show how they can be constructed by taking suitable derivatives of a light-like Wilson loop in twistor space and shrinking it down to a point.
In particular, these vertices directly yield the tree-level MHV super form factors of all composite operators in $\mathcal{N}=4$ SYM theory. 


\newpage

\setcounter{tocdepth}{2}
\hrule height 0.75pt
\tableofcontents
\vspace{0.8cm}
\hrule height 0.75pt
\vspace{1cm}

\setcounter{tocdepth}{2}

\section{Introduction}
Arguably the simplest interacting gauge theory in four dimensions is $\mathcal{N}=4$ Super Yang-Mills (SYM). Studying this theory has led to an explosion of important theoretical developments, such as holography (AdS/CFT) 
and integrability in the planar limit, see \cite{Beisert:2010jr} for a review of the latter. In parallel, it has helped the development of on-shell techniques, reviewed in \cite{Elvang:2013cua, Henn:2014yza}, which are extremely efficient for the computation of amplitudes.

In addition, the action of $\calN =4$ SYM admits a formulation in twistor space \cite{Boels:2006ir} in terms of a single superfield $\calA$; see \cite{Adamo:2011pv} for a review. In contrast to the usual space-time action, which contains only cubic and quartic vertices, the twistor action is a perturbation around the self-dual point of the theory and contains an infinite sum of interaction vertices of increasing valency. At first glance, this might seem to complicate things. However, choosing a specific axial gauge and inserting on-shell external states directly into these vertices, one immediately obtains all tree-level MHV (super) amplitudes \cite{Boels:2007qn}, which shows that the twistor action does repackage the interactions in a more economical way. More generic N$^{k}$MHV amplitudes are then built from $k+1$ of these vertices and $k$ twistor-space propagators \cite{Adamo:2011cb,Adamo:2011pv}. Furthermore, it was shown in \cite{Mason:2010yk} and \cite{Bullimore:2011ni} that the integrand of a planar scattering amplitude corresponds to the planar integrand of a supersymmetric Wilson loop in twistor space. 
However, not only has the twistor-space formalism been successfully employed for the computation of amplitudes \cite{ArkaniHamed:2009dn,Mason:2009qx,Bullimore:2010pj,Adamo:2011cb,Adamo:2011pv} and the dual light-like Wilson loop expectation values\cite{CaronHuot:2010ek, Mason:2010yk,Belitsky:2011zm}, but recently \cite{Koster:2014fva, Chicherin:2014uca} it was also applied for the computation of correlation functions of certain gauge-invariant local composite operators (or composite operators, for short). 

Dealing with composite operators $\calO$ in twistor space, however, is subtle \cite{Koster:2016ebi}. Since a point $x$ in Minkowski space corresponds to a line in twistor space, gauge invariance requires that the component fields of $\calO$, placed on different twistors on the line, be connected by the parallel propagator $U$ for the gauge connection $\calA$. Given that $U$ is a path-ordered exponential of $\calA$, it contains infinitely many terms. This turns out to be a simplification rather than a complication for the computation of the MHV form factors of the composite operator, since expanding $U$ in terms of $\calA$ directly generates all the terms needed for the emission of an arbitrary number of positive helicity gluons. 
Hence, twistor space does for form factors exactly what it also does for amplitudes, namely package the MHV objects neatly into a simple expression. 
However, in order to obtain all interactions of the operators at MHV level, e.g.\ also the splitting of a scalar into two anti-fermions, these terms have to be included in the twistor-space description of the composite operators as well.

 Let us now introduce the form factors in more detail. They can be thought of as a bridge between the on-shell amplitudes and the off-shell correlation functions.
Letting $\fp_i$ with $\fp_i^2=0$ be the momenta of the on-shell states, and $\fq$ with $\fq^2\neq0$ the momentum of the operator $\calO$, the form factor is defined as the expectation value of $\calO$ with $n$ on-shell states $\Phi_i$:
\begin{equation}
\label{eq:formfactordefinition}
\calF_{\calO}(1^{\Phi_1},\ldots, n^{\Phi_n};\fq)= \int \frac{\dd^4x}{(2\pi)^4}\, \e^{-i\fq x}\melement{\Phi_1(\fp_1)\cdots \Phi_n(\fp_n)}{\calO(x)}{0}\,.
\end{equation}
Form factors in $\mathcal{N}=4$ SYM have received increasing attention in recent years, both at weak coupling \cite{vanNeerven:1985ja,Brandhuber:2010ad,Bork:2010wf,Brandhuber:2011tv,
Bork:2011cj,Henn:2011by,Gehrmann:2011xn,Brandhuber:2012vm,Bork:2012tt,
Engelund:2012re,Johansson:2012zv,Boels:2012ew,Penante:2014sza,
Brandhuber:2014ica,Bork:2014eqa,Wilhelm:2014qua,Nandan:2014oga,Loebbert:2015ova,
Bork:2015fla,Frassek:2015rka,Boels:2015yna,Huang:2016bmv} and at strong coupling \cite{Alday:2007he,Maldacena:2010kp,Gao:2013dza}; see \cite{Wilhelm:2016izi} for a review. In comparison to amplitudes, however, where all tree-level expressions \cite{Drummond:2008cr} as well as the unregularized integrands of all loop-level expressions \cite{ArkaniHamed:2010kv} have been found, much less is known for form factors. All tree-level form factors are known only for the operators of the stress-tensor supermultiplet \cite{Bork:2014eqa}. Apart from that, expressions for $n$-point form factors exist only at MHV level for operators of the $SU(2)$ subsector and twist-two operators of the $SL(2)$ subsectors \cite{Engelund:2012re}. Tree-level form factors for general composite operators are known only for the minimal non-trivial particle multiplicity \cite{Wilhelm:2014qua}.

The goal of this article is to systematize and extend the framework introduced in \cite{Koster:2016ebi} to describe all composite operators of $\calN=4$ SYM in twistor space. Then, upon inserting external on-shell particles, we produce all tree-level MHV (super) form factors of the theory. We start by reviewing the construction of composite operators in space-time in section \ref{sec:review}. 
Then, in section \ref{sec:construction}, we extend the idea proposed in \cite{Koster:2016ebi} of the Wilson loop as a generating object for the operator vertex to all composite operators.  
Finally, in section \ref{sec:MHVformfactors}, we insert momentum eigenstates into these operator vertices found from the Wilson loop construction. In this way, we derive closed expressions first for all minimal tree-level MHV form factors and then also for all $n$-point tree-level MHV form factors. 
Note that throughout this paper we consider all amplitudes and form factors to be color-ordered and super.
We conclude in section \ref{sec:conclusion} with a brief outlook on future research, while 
three appendices contain details of the Wilson loop in our construction, instructive examples of operator vertices and the detailed derivation of the $n$-point MHV form factors.

\section{Twistor space and composite operators in space-time}
\label{sec:review}

Before constructing the twistor-space vertices for all composite operators and computing their MHV form factors, let us first review how the twistor action yields all MHV amplitudes in subsection \ref{subsec: twistors and amplitudes}. We then recall the form of the composite operators in space-time in subsection \ref{subsec: composite operators}.

\subsection{Amplitudes in twistor space}
\label{subsec: twistors and amplitudes}

We start with chiral Minkowski superspace $\mathbb{M}^{4|8}$, in which eight Gra\ss mann variables $\theta^{\alpha a}$ are attached to each point $x^{\alpha\dot\alpha}$ in Minkowski space with $\alpha, \dot{\alpha}\in \{1,2\}$, $a\in\{1,2,3,4\}$. In parallel, we have supertwistor space $\mathbb{CP}^{3|4}$, whose elements are called supertwistors and are written as $\calZ = (\la_{\alpha},\mu^{\dot\alpha},\chi^a)$, where the $\chi^a$ are Gra\ss mann parameters. 
Each point $(x,\theta)$ in $\mathbb{M}^{4|8}$ corresponds to a unique projective line in supertwistor space  given by the set of supertwistors 
\begin{equation}
\label{eq:definitioncalZonaline}
\calZ_x(\lambda)=(\la_{\alpha},ix^{\alpha\dot\alpha}\la_{\alpha},i\theta^{a\alpha}\la_{\alpha})\,,\quad \lambda\in \mathbb{CP}^1\,.
\end{equation}
For brevity, we will denote this line in $\mathbb{CP}^{3|4}$ by $x$ instead of $(x,\theta)$. 

The  twistor action is written using a single connection superfield $\calA$. This superfield is expanded in the Gra\ss mann parameters $\chi^a$ as
\beq
\label{eq:expansionAAA}
\AAA(\mathcal{Z})=g^+ +\chi^a\bar{\psi}_a+\frac{1}{2}\chi^a\chi^b\phi_{ab}+\frac{1}{3!}\chi^a\chi^b\chi^c\psi^d\epsilon_{abcd}+\chi^1\chi^2\chi^3\chi^4 g^-\eqncom
\eeq
where the components fields $g^\pm,\ldots$ do not depend on $\chi$.
The twistor action is the sum of two parts $\calS_1+\calS_2$, where $\calS_1$, introduced in  \cite{Witten:2003nn}, describes the self-dual part and $\calS_2$ is referred to as the interaction part \cite{Boels:2007qn}.
In a specific axial gauge, called the CSW gauge, the cubic term in $\calS_1$ vanishes and all interactions indeed come from the $\calS_2$ part of the action.
Expanding $\calS_2$ gives
\begin{equation}
\label{eq:interactionexpanded}
\calS_2=-\frac{g_\YM^2}{4}\int_{\mathbb{M}^{4|8}} \frac{\dd^4 z\:\dd^8\theta }{(2\pi)^4} \sum_{n=1}^{\infty}\frac{1}{n} \int_{(\mathbb{CP}^1)^n}\DD\la_1\DD\la_2\cdots \DD\la_n\frac{\Tr\big(\AAA(\la_1)\cdots\AAA(\la_n)\big)}{ \abra{\lambda_1}{\lambda_2} \cdots \abra{\lambda_{n-1}}{ \lambda_{n}} \abra{\lambda_{n}}{\lambda_{1}}}
\eqncom
\end{equation}
where $\abra{\la_i}{\la_j} =\epsilon^{\alpha\beta}\la_{i\alpha} \la_{j\beta}=\la_{i}^\alpha \la_{j\alpha} $, $\DD\la_i\equiv\frac{\abra{\la_i}{\dd\la_i}}{2\pi i}$ and $\AAA(\la_i)\equiv \AAA(\calZ_{z}(\la_i))$. In our conventions, $\epsilon^{\alpha\beta}=\epsilon^{\dot\alpha\dot\beta}$ and $\epsilon^{12}=\epsilon_{21}=1$.

As every propagator increases the MHV degree by one, the $n$-point MHV amplitudes are directly obtained from the corresponding summands in \eqref{eq:interactionexpanded}, see \cite{Boels:2007qn}.
In order to obtain these amplitudes, one simply inserts external on-shell momentum states. In twistor language, they are written as 
\begin{equation}
 \label{eq:definitiononshellmomentumeigenstates}
\calA_{\fP}(\calZ)=2\pi i \int_{\mathbb{C}}\frac{\dd s}{s}\e^{s(\sbra{\bar{p}}{\mu}+\{\chi\eta\})}
\bar{\delta}^2(s\lambda_{\alpha}-p_{\alpha})\eqnsem
\end{equation}
see \cite{Cachazo:2004kj,Adamo:2011cb}. In \eqref{eq:definitiononshellmomentumeigenstates}, we used $\sbra{\mu}{\mu'}=\epsilon_{\dot \alpha\dot\beta}\mu^{\dot \alpha}{\mu'}^{\dot \beta}=\mu_{\dot \alpha}{\mu'}^{\dot\alpha}$, $\{\chi\eta\}=\chi^a\eta_a$
as well as $\bar{\delta}^2(\lambda)= \bar{\delta}^1(\lambda_1)\bar{\delta}^1(\lambda_2)$ with $\bar{\delta}^1(z)=\frac{1}{2\pi i}\dd z\,\bar\partial \big(\frac{1}{z}\big)$ being the $\delta$ function on the complex plane. 
We denote the on-shell supermomenta in terms of super-spinor-helicity variables as $\fP=(\fp_{\alpha\dot{\alpha}},\eta_a)\equiv (p_{\alpha},\bar{p}_{\dot{\alpha}},\eta_a)$.%
\footnote{Note that $\lambda_{\alpha}\in\mathbb{CP}^1$ is a coordinate in position twistor space that is integrated over in each vertex, while $p_{\alpha}$ is one of the spinor-helicity variables of momentum space. Although being conceptually different quantities, their values are being related when inserting the momentum eigenstate \eqref{eq:definitiononshellmomentumeigenstates} as discussed below.}
Inserting the on-shell states \eqref{eq:definitiononshellmomentumeigenstates} into the $n$\textsuperscript{th} summand of \eqref{eq:interactionexpanded} and taking into account the $n$ ways of cyclically attaching them yields the tree-level $n$-point MHV amplitude:
\begin{equation}
\label{eq:MHV amplitude}
\begin{split}
\mathscr{A}^{\text{MHV}}(\fP_1, \dots , \fP_n) &= n\int\frac{\dd^4z\dd^8\theta}{(2\pi)^4}\frac{1}{n}\int \DD\lambda_1\cdots \DD\lambda_n\frac{\calA_{\fP_1}(\calZ_z(\lambda_1))\cdots \calA_{\fP_1}(\calZ_z(\lambda_n))}{\abra{\lambda_1}{\lambda_n}\cdots \abra{\lambda_n}{\lambda_1}}\\
&=\int\frac{\dd^4z\dd^8\theta}{(2\pi)^4}\frac{\e^{iz\sum_{j=1}^np_j\tilde{p}_j+i\theta\sum_{j=1}^np_j\eta_j}}{\abra{1}{2}\cdots \abra{(n-1)}{n}\abra{n}{1}}=\frac{\delta^{4|8}(\sum_{i=1}^n \fP_i)}{\prod_{k=1}^n\abra{k}{(k+1)}} \eqncom
\end{split}
\end{equation}
with $\abra{i}{j}\equiv \abra{p_i}{p_j}$. We remark that the integrations over $s_k$ from the definition of $\calA_{\fP_k}(\calZ_z(\la_k))$ as well as over its corresponding $\la_k$ effectively cancels $s_k$ and replaces $\la_k\rightarrow p_k$, $\mu_k^{\dot{\alpha}}\rightarrow iz^{\alpha \dot{\alpha}}p_{k,\alpha}$ and $\chi_k^a\rightarrow i\theta^{\alpha a }p_{k,\alpha}$ due to the $\bar{\delta}^2$ function and the parametrization \eqref{eq:definitioncalZonaline}. 
Here and in the rest of this paper, we have moreover dropped the explicit specification of the integration range.  

In this paper, we find a similar result for the tree-level MHV form factors of all composite operators.

\subsection{Composite operators}
\label{subsec: composite operators}

Let us now explain how composite operators are constructed in Minkowski space-time; see e.g.\ \cite{Beisert:2004ry,Minahan:2010js} for reviews.
In the planar limit, it is sufficient to look at single-trace operators $\mathcal{O}$. Denoting the covariant derivative  by $D$ and suppressing indices for the time being, these operators are constructed by tracing over products of covariantly transforming fields $D^{k_\ri}A_\ri$ with $k_i=0,1,2,\ldots$ placed at the same space-time point $x$:
\begin{equation}
\label{eq:composite operator}
\mathcal{O}(x)=\Tr\left(D^{k_1}A_1(x)D^{k_2}A_2(x)\cdots D^{k_L}A_L(x)\right)\eqndot
\end{equation}
The  fields $A_\ri$ are taken from the set of the six scalars $\phi_{ab}$, the four fermions $(\psi_{bcd})_{\alpha}=\epsilon_{abcd}\psi^d_{\alpha}$, the four anti-fermions $\bar{\psi}_{a\dot{\alpha}}$ and the field strength $F_{\mu\nu}$.
By contracting with the Pauli matrices $\sigma^{\mu}_{\alpha\dot{\alpha}}$, we can get rid of the $\mu, \nu$ indices and will henceforth only use the spinor ones $\alpha$ and $\dot{\alpha}$.
The field strength is thus split into a self-dual and an anti-self-dual part as
\begin{equation}
F_{\mu\nu}(\sigma^\mu)_{\alpha\dot{\alpha}}(\sigma^\nu)_{\beta\dot{\beta}}\propto \epsilon_{\dot{\alpha}\dot{\beta}}F_{\alpha\beta}+\epsilon_{\alpha\beta}\bar{F}_{\dot{\alpha}\dot{\beta}} \eqndot
\end{equation}
When considering covariant derivatives
$D_{\alpha\dot{\alpha}}$
that act on the fields, we use the equations of motion of the fields, the definition of the field strength and the Bianchi identity for the field strength to replace any antisymmetric combination of the spinor indices $\alpha$, $\dot{\alpha}$ by a product of terms that are individually fully symmetric in the spinor indices. 
All composite operators can therefore be built from the following set of fields:
\begin{align}
\label{eq: alphabet of fields}
 D^kA\in\{ 
 &D_{(\alpha_1\dot\alpha_1}\cdots D_{\alpha_k\dot\alpha_k}\bar{F}_{\dot\alpha_{k+1}\dot\alpha_{k+2})},&
 &D_{(\alpha_1\dot\alpha_1}\cdots D_{\alpha_k\dot\alpha_k}\bar{\psi}_{\dot\alpha_{k+1})a},& 
 &D_{(\alpha_1\dot\alpha_1}\cdots D_{\alpha_k\dot\alpha_k)}\phi_{ab},&\notag\\
 &D_{(\alpha_1\dot\alpha_1}\cdots D_{\alpha_k\dot\alpha_k}\psi_{\alpha_{k+1})abc},&
 &D_{(\alpha_1\dot\alpha_1}\cdots D_{\alpha_k\dot\alpha_k}F_{\alpha_{k+1}\alpha_{k+2})} \}\eqncom & &&
\end{align}
where the parentheses denote symmetrization in the spinor indices $\alpha$, $\dot{\alpha}$ and the fields are antisymmetric in the flavor indices $a$.
These fields form an irreducible representation of the symmetry algebra $\mathfrak{psu}(2,2|4)$, the so-called singleton representation $\textsf{V}_{\textsf{S}}$. 
They form the spin chain of $\mathcal{N}=4$ SYM.

In practice, the explicit symmetrization of the spinor indices can be quite cumbersome. This can be circumvented by introducing a 
light-like polarization vector $\ftau^{\alpha\dot\alpha}=\tau^{\alpha}\bar{\tau}^{\dot\alpha}$. 
Moreover, we include a Gra\ss mann variable $\xi^a$ which combines with  $\ftau$ to form the  superpolarization vector  $\fTau=(\tau^{\alpha},\bar{\tau}^{\dot\alpha}, \xi^a)$.
We can then write the fields in \eqref{eq: alphabet of fields} as
\begin{equation}
\label{eq: alphabet of fields with polarizations}
\begin{aligned}
 D^kA\in\{ 
 &+\tau^{\alpha_1}\dots\tau^{\alpha_{k\phantom{+1}}}\bar{\tau}^{\dot\alpha_1}\dots\bar{\tau}^{\dot\alpha_{k+2}}\phantom{\xi^a}\phantom{\xi^a}\phantom{\xi^a}\phantom{\xi^a}
 D_{\alpha_1\dot\alpha_1}\cdots D_{\alpha_k\dot\alpha_k}\bar{F}_{\dot\alpha_{k+1}\dot\alpha_{k+2}},\\
 &+\tau^{\alpha_1}\dots\tau^{\alpha_{k\phantom{+1}}}\bar{\tau}^{\dot\alpha_1}\dots\bar{\tau}^{\dot\alpha_{k+1}}\xi^a\phantom{\xi^a}\phantom{\xi^a}\phantom{\xi^a}
 D_{\alpha_1\dot\alpha_1}\cdots D_{\alpha_k\dot\alpha_k}\bar{\psi}_{\dot\alpha_{k+1}a},\\
 &-\tau^{\alpha_1}\dots\tau^{\alpha_{k\phantom{+1}}}\bar{\tau}^{\dot\alpha_1}\dots\bar{\tau}^{\dot\alpha_{k\phantom{+1}}}\xi^a\xi^b\phantom{\xi^a}\phantom{\xi^a}
 D_{\alpha_1\dot\alpha_1}\cdots D_{\alpha_k\dot\alpha_k}\phi_{ab},\\
 &-\tau^{\alpha_1}\dots\tau^{\alpha_{k+1}}\bar{\tau}^{\dot\alpha_1}\dots\bar{\tau}^{\dot\alpha_{k\phantom{+1}}}\xi^a\xi^b\xi^c\phantom{\xi^a}
 D_{\alpha_1\dot\alpha_1}\cdots D_{\alpha_k\dot\alpha_k}\psi_{\alpha_{k+1}abc},\\
 &+\tau^{\alpha_1}\dots\tau^{\alpha_{k+2}}\bar{\tau}^{\dot\alpha_1}\dots\bar{\tau}^{\dot\alpha_{k\phantom{+1}}}\xi^a\xi^b\xi^c\xi^d
 D_{\alpha_1\dot\alpha_1}\cdots D_{\alpha_k\dot\alpha_k}F_{\alpha_{k+1}\alpha_{k+2}abcd}
 \}\eqncom
\end{aligned}
\end{equation}
where $F_{\alpha_{k+1}\alpha_{k+2}abcd}=\frac{1}{4!}\epsilon_{abcd}F_{\alpha_{k+1}\alpha_{k+2}}$.
Note that an independent superpolarization vector can be chosen for each field $D^{k_\ri}A_\ri$ of \eqref{eq:composite operator}, i.e.\ at each site of the spin chain. 
The expressions \eqref{eq: alphabet of fields} can be recovered by taking suitable derivatives of \eqref{eq: alphabet of fields with polarizations}
with respect to the superpolarization vector.
For example,
\begin{equation}
\label{eq: specifying components by derivatives}
 \begin{aligned}
  D_{(\alpha_1\dot\alpha_1}\cdots D_{\alpha_k\dot\alpha_k)}\phi_{ab}
  &=
  \frac{1}{k! k! 2!}
  \frac{\partial}{\partial \tau^{\alpha_1}}\cdots\frac{\partial}{\partial \tau^{\alpha_k}} 
  \frac{\partial}{\partial \bar{\tau}^{\dot\alpha_1}}\cdots\frac{\partial}{\partial \bar{\tau}^{\dot\alpha_k}} 
  \frac{\partial}{\partial \xi^{a}}\frac{\partial}{\partial \xi^{b}} \\
  &\phaneq(-1)\tau^{\beta_1}\dots\tau^{\beta_{k}}\bar{\tau}^{\dot\beta_1}\dots\bar{\tau}^{\dot\beta_{k}}\xi^c\xi^d
 D_{\beta_1\dot\beta_1}\cdots D_{\beta_k\dot\beta_k}\phi_{cd}
  \eqndot
 \end{aligned}
\end{equation}
We will see in the next section how the operators built from \eqref{eq: alphabet of fields with polarizations} can be constructed using the twistor-space formalism.

\section{Construction of composite operators in twistor space}
\label{sec:construction}
In this section, we construct space-time-local composite operators from non-local ones in twistor space. Our method is very physical and involves the shrinking of suitably decorated Wilson loops. The underlying rationale is that composite operators are traces of products of the irreducible fields \eqref{eq: alphabet of fields with polarizations}, which we imagine being linked by (infinitesimal) flux lines. We have already communicated the resulting expression for the scalar field $\phi_{ab}$ in twistor space in \cite{Koster:2016ebi}. In this paper, we give considerable additional detail for our construction. Furthermore, we derive the vertices for all irreducible fields \eqref{eq: alphabet of fields with polarizations}, and hence for all composite operators.
Concretely, we obtain these by applying derivatives to a Wilson loop in twistor space and taking the limit where the Wilson loop becomes a point in Minkowski space, i.e.\ a line in twistor space.

\subsection{Wilson loops in twistor space}
\label{subsec:Wilsonloopsintwistorspace}
Let us start by constructing a Wilson loop in twistor space. For this, we take 
a set of $n$ points in  Minkowski superspace $\{(x_i,\theta_i)\}_{i=1}^n$ such that $x_{i-1}$ and $x_{i}$ are light-like separated for all $i$ and $x_{n}=x_{0}$.
Due to this light-like separation, any pair of the associated successive lines $x_{i-1}$ and $x_{i}$ in twistor space intersects in a twistor, which we denote by $\calZ_{i}$. 
\begin{figure}[htbp]
 \centering
  \includegraphics[height=3.5cm]{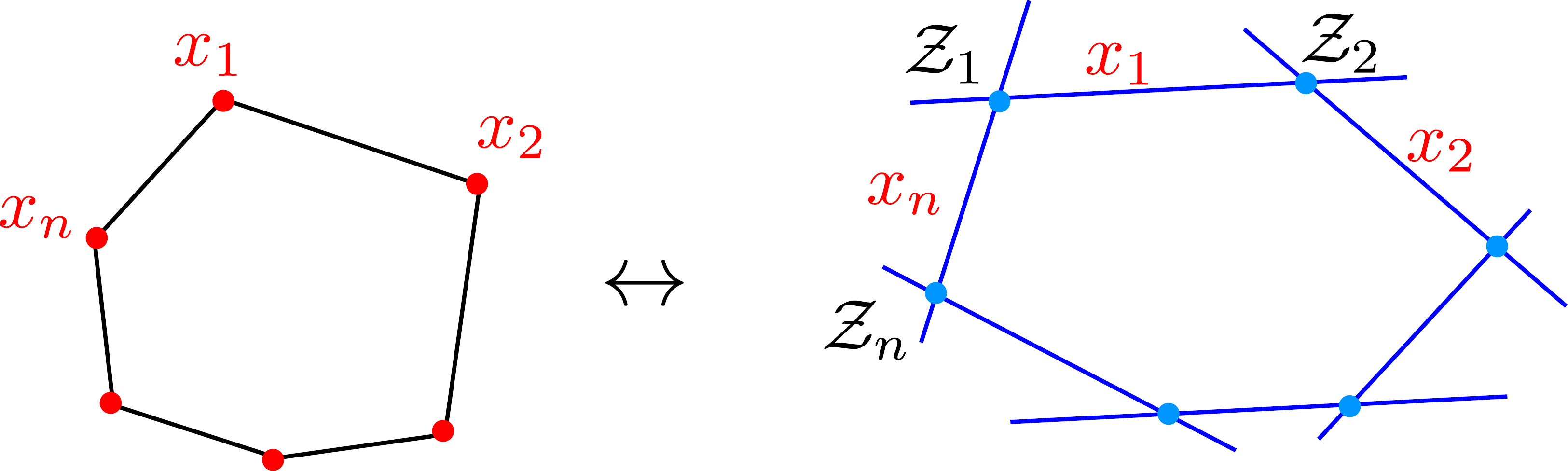}
  \caption{\it A supersymmetric $n$-gonal Wilson loop in position space and its twistor space analogue.
  }
  \label{fig:LgonalWilson}
\end{figure}
It turns out that constructing an appropriate light-like Wilson loop that will satisfy our requirements is not straightforward. The precise geometry of the 
Wilson loop we will use will be hinted at later in subsection \ref{subsec:our wilson loop construction} and is explained in detail in appendix \ref{app:geometry}. 
For our current purpose, however, it is sufficient to consider the simplified construction and notation; see figure~\ref{fig:LgonalWilson}. It is written as 
\begin{align}
\label{eq: polygonal Wilson loop}
& \WWW(x_1,\ldots, x_n)= \Tr\left( U_{x_1}(\calZ_1,\calZ_2) U_{x_2}(\calZ_2,\calZ_3) \cdots U_{x_n}(\calZ_n,\calZ_1)   \right)\eqncom
\end{align}
where $U_{x_i}(\calZ_i,\calZ_{i+1})$ is the parallel propagator connecting the supertwistors $\calZ_i$ and $\calZ_{i+1}$ on the line $x_i$ in $\mathbb{CP}^{3|4}$. Its expression is given by a path-ordered exponential \cite{Bullimore:2011ni}, which can be expanded as
\begin{equation}
\label{eq:frameUdefinitionpart1}
 U_{x_{i}}(\calZ_i,\calZ_{i+1})=1+\sum_{m=1}^\infty \int_{}\frac{\abra{\lambda_i}{\lambda_{i+1}}\DD\tilde{\la}_1 \cdots  \DD\tilde{\la}_m}{\abra{\lambda_i}{\tilde{\la}_1}\abra{\tilde{\la}_1}{\tilde{\la}_2}\cdots \abra{\tilde{\la}_m}{\lambda_{i+1}}}\AAA(\calZ_{x_i}(\tilde{\la}_1)) \cdots \AAA(\calZ_{x_i}(\tilde{\la}_m))
\eqndot
\end{equation}
We often write $U_{x_i}(\la_i,\la_{i+1}) \equiv U_{x_i}(\calZ_i,\calZ_{i+1})$, where it is understood that $\calZ_i=\calZ_{x_i}(\la_i)$ and $\calZ_{i+1}=\calZ_{x_i}(\la_{i+1})$.


\subsection{From Wilson loops to local composite operators}
\label{subsec:compositeoperatorss}

Let us now explain how to obtain the vertices for composite operators in space-time from the Wilson loops \eqref{eq: polygonal Wilson loop}. 
The crucial idea is to act on the Wilson loop with derivative operators with respect to the superspace coordinates $x_i$ and $\theta_i$.

For the current discussion, it is sufficient to concentrate on a specific edge of the loop, i.e.\ we just look at only one of the parallel propagators $U_{x_{\ri}}(\la_\ri,\la'_\ri)$ of \eqref{eq: polygonal Wilson loop}. In anticipation of notation to be introduced in appendix \ref{app:geometry}, we write $\la_i'=\la_{i+1}$.
Taking a derivative with respect to the superspace coordinates $x_{\ri}$ and $\theta_{\ri}$ of a parallel propagator $U_{x_{\ri}}$ inserts a twistor field $\AAA$ in the middle of it. To be more precise, acting on the parallel propagator with the derivative with respect to $\theta^{\alpha a}_{\ri}$ gives \cite{Adamo:2011dq}
\begin{equation}
\label{derivativeonUU}
\frac{\partial}{\partial \theta_{\ri}^{\alpha a}}U_{x_{\ri}}(\la_{\ri},\la'_{\ri})=i\int \frac{\DD\tilde{\la} \abra{\la_{\ri}}{\la'_{\ri}}}{\abra{\la_{\ri}}{\tilde{\la}}\abra{ \tilde{\la}}{\la'_{\ri}}} U_{x_{\ri}}(\la_{\ri},\tilde{\la})\tilde{\la}_{\alpha}\frac{\partial \calA(\tilde\la)}{\partial {\tilde{\chi}}^a}U_{x_{\ri}}(\tilde{\la},\la'_{\ri})\eqncom
\end{equation}
while acting with its bosonic sibling leads to%
\footnote{Strictly speaking, \eqref{eq: dU/dx} is incomplete. For an off-shell field, the derivative obtains an extra term proportional to $\bar\mu$, the conjugate of $\mu$. However, this term vanishes when one inserts an on-shell external state and can therefore be discarded. We thank Lionel Mason for discussion on this point.}
\begin{equation}
\label{eq: dU/dx}
	\frac{\partial}{\partial x_{\ri}^{\alpha \dot{\alpha}}}U_{x_{\ri}}(\la_{\ri},\la'_{\ri})=i\int \frac{  \DD\tilde{\la}\abra{\la_{\ri}}{\la'_{\ri}}}{\abra{\la_{\ri}}{\tilde{\la}}\abra{ \tilde{\la}}{\la'_{\ri}}} U_{x_{\ri}}(\la_{\ri},\tilde{\la})\tilde{\la}_{\alpha}\frac{\partial \calA(\tilde{\la})}{\partial {\tilde{\mu}}^{\dot{\alpha}}}U_{x_{\ri}}(\tilde{\la},\la'_{\ri})\eqndot
\end{equation}

\begin{figure}[htbp]
 \centering
  \includegraphics[height=3.5cm]{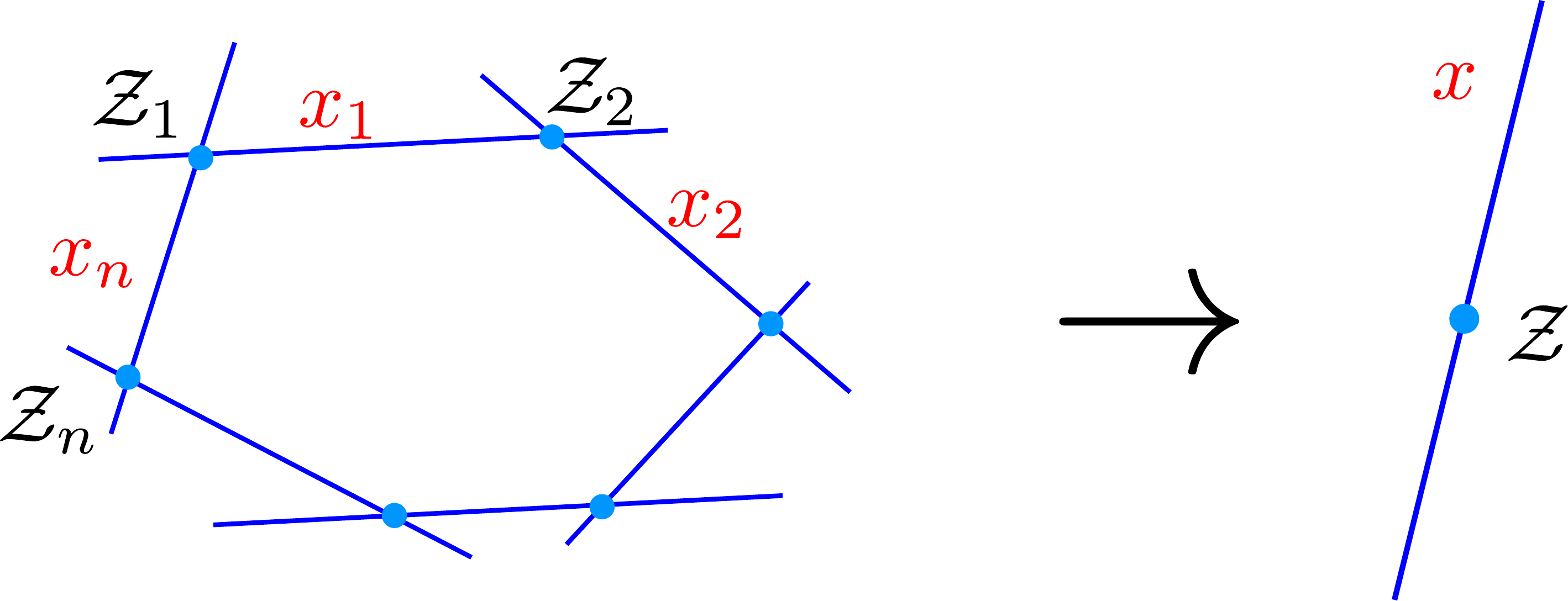}
  \caption{\it This figure sketches the limit procedure which sends the light-like Wilson loop to a point $x$, or in twistor space, to a line. The details are considerably more complicated, see appendix \ref{app:geometry}. 
  }
  \label{fig:operatorlimit}
\end{figure}

We now demonstrate how to produce the field content \eqref{eq: alphabet of fields with polarizations} of $\mathcal{N}=4$ SYM from the above derivatives, using the scalar field as an example. 
In order to obtain the operator vertex of \cite{Koster:2016ebi} for $\phi=-\xi^a\xi^b\phi_{ab}$ from the parallel propagator, we first act with two derivative operators. This can be easily shown to yield the expression
\begin{align}
\label{eq:doublethetaderivativeofU}
&\frac{\partial^2U_{x_{\ri}}(\la_{\ri},\la'_{\ri})}{\partial \theta_{\ri}^{\alpha a}\partial \theta_{\ri}^{\beta b}}
=i^2\!\int
\DD\tilde{\la}_1\frac{\abra{\la_{\ri}}{\la'_{\ri}}}{\abra{\la_{\ri}}{\tilde{\la}_1}\abra{\tilde{\la}_1}{\la'_{\ri}}}U_{x_{\ri}}(\la_{\ri},\tilde{\la}_1)\tilde{\la}_{1\alpha}\tilde{\la}_{1\beta}\frac{\partial^2\calA(\tilde{\la}_1)}{\partial {\tilde{\chi}_1}^a\partial {\tilde{\chi}_1}^b}U_{x_{\ri}}(\tilde{\la}_1,\la'_{\ri})\\
&+i^2\!\int
\frac{\DD\tilde{\la}_1 \DD\tilde{\la}_2\abra{\la_{\ri}}{\la'_{\ri}}}{\abra{\lambda}{\tilde{\la}_1}\abra{\tilde{\la}_1}{\tilde{\la}_2}\abra{\tilde{\la}_2}{\la'_{\ri}}}U_{x_{\ri}}(\la_{\ri},\tilde{\la}_1)\tilde{\la}_{1\alpha}\frac{\partial\calA(\tilde{\la}_1)}{\partial{\tilde{\chi}_1}^a}U_{x_{\ri}}(\tilde{\la}_1,\tilde{\la}_2)\tilde{\la}_{2\beta}\frac{\partial\calA(\tilde{\la}_2)}{\partial{\tilde{\chi}_2}^b}U_{x_{\ri}}(\tilde{\la}_2,\la'_{\ri})-\binom{\alpha\leftrightarrow\beta}{a\leftrightarrow b}\,.\nonumber
\end{align}
To extract the scalar field $\phi$ from this, we need to get rid of the $\alpha$, $\beta$ indices and also keep the expression homogeneous in $\la_{\ri}$ and $\la'_{\ri}$ of degree $0$. This can be achieved by contracting \eqref{eq:doublethetaderivativeofU} with 
\begin{equation}
\label{eq: contraction prefactor}
\frac{\la_{\ri}^{\alpha}{\la'}_{\ri}^{\beta}}{\langle\la_{\ri}\la'_{\ri}\rangle} \eqndot
\end{equation}
We also contract with the fermionic polarization vectors $-\xi_{\ri}^a\xi_{\ri}^b$ to remove the Gra\ss mann indices. Furthermore, as we want to use these fields to construct \textit{local} composite operators, we have to take the limit where all space-time points go to the same point $x$, see figure \ref{fig:operatorlimit}. 
The resulting expression should only depend on the point $x$ in space-time and not on the geometry of the Wilson loop that we used to construct it. In particular, it should not depend on the coordinates $\la_{\ri}$ and $\la'_{\ri}$ of the corners of the loop. This independence can be achieved by taking the limit $\la_{\ri}\parallel\la'_{\ri}$, and we choose the normalization such that $\la'_{\ri}\rightarrow\la_{\ri}$. However, this comes with a potential problem. Specifically, suppose we act with derivatives on each edge of the loop of figure~\ref{fig:LgonalWilson} and then take the limit such that $\la_\ri\rightarrow \la_{\ri+1}$ for each $\ri$. Then, in this limit, the geometry would be somewhat ill-defined, since at least two different twistors are needed to define a line in twistor space. To go around this issue, we \textit{add extra edges} to the loop, which we describe in full detail in section \ref{subsec:our wilson loop construction} and appendix \ref{app:geometry}. 

We call the limit of shrinking the Wilson loop to a point and $\la_{\ri}\rightarrow \la'_{\ri}$ \textit{the operator limit} and write it as $\hexagon\rightarrow \xdot$. 
Taking this limit, we find that%
\footnote{In the third term of \eqref{eq:doublethetaderivativeofU}, we used the Schouten identity and the Gra\ss{}mannian nature of $\xi_{\ri}^a$, $\xi_{\ri}^b$ to find the same result as from the second term.}
\begin{align}
\label{scalar}
&\lim_{\hexagon\rightarrow \xdot}\Bigg[-\frac{\la_{\ri}^{\alpha}{\la'_{\ri}}^{\beta}}{\langle\la_{\ri}\la'_{\ri}\rangle}\xi_{\ri}^a\xi_{\ri}^b\frac{\partial^2U_{x_{\ri}}(\la_{\ri},\la'_{\ri})}{\partial \theta_{\ri}^{\alpha a}\partial \theta_{\ri}^{\beta b}}\Bigg]_{\big{|}\theta=0}
=-\int \DD\tilde{\la}_1\, U_{x}(\la_i,\tilde{\la}_1) \xi_{\ri}^a\xi_{\ri}^b\frac{\partial^2 \AAA(\tilde{\la}_1)}{\partial\tilde{\chi}_1^a\partial\tilde{\chi}_1^b}U_{x}(\tilde{\la}_1,\la_i){}_{\big{|}\theta=0} \nonumber\\
 &\qquad  \phaneq
+2\int \frac{\DD\tilde{\la}_1 \DD\tilde{\la}_2}{\sabra{\tilde{\la}_1}{\tilde{\la}_2}} U_{x}(\la_i,\tilde{\la}_1) \xi_{\ri}^a\frac{\partial \AAA(\tilde{\la}_1)}{\partial\tilde\chi_1^a}U_x(\tilde{\la}_1,\tilde{\la}_2)\xi_{\ri}^b\frac{\partial \AAA(\tilde{\la}_2)}{\partial\tilde{\chi}_2^{b}}U_{x}(\tilde{\la}_2,\la_i)_{\big{|}\theta=0}\\
 &\qquad =  h_x(\la_i)\textbf{W}_{\phi(x)} h_x(\la_i)^{-1} \eqncom\nonumber
\end{align}
where we set $\theta=0$ (this sets the Gra\ss mann parameters $\chi$ to zero after taking the derivatives, because we are only interested in the $\phi_{ab}$ component) and have defined
\begin{equation}
\label{eq: vertex scalar}
\begin{aligned}
\textbf{W}_{\phi(x)}&=- 
 \int \DD\la h^{-1}_x(\la) \xi_{\ri}^a\xi_{\ri}^b\frac{\partial^{2} \AAA(\la)}{\partial\chi^{a}\partial\chi^{b}}h_x(\la)_{\big{|}\theta=0}
  \\
 &\phaneq
+2\int \DD\la\DD\la' h^{-1}_x(\la) \xi_{\ri}^a\frac{\partial \AAA(\la)}{\partial\chi^{a}}
\frac{U_x(\la,\la')}{\abra{\la}{\la'}}\xi_{\ri}^b\frac{\partial \AAA(\la')}{\partial\chi'^{b}}h_x(\la')_{\big{|}\theta=0} \eqndot 
\end{aligned}
\end{equation}
Here, $h_x(\la)$ is the frame on $x$  that trivializes the connection $\calA$ along the line $x$ and hence ensures that traces of products of \eqref{scalar} are gauge invariant  \cite{Adamo:2011dq}. 
It is related to the parallel propagator $U_x$ as
\begin{equation}
\label{eq: parallel propagator in terms of frames}
 U_x(\la,\tilde\la)=h_x(\la)h_x(\tilde\la)^{-1} \eqndot
\end{equation}
The expression $\textbf{W}_{\phi(x)}$ in \eqref{eq: vertex scalar} perfectly agrees with the proposed vertex $\textbf{W}_{\phi_{ab}(x)}$ of \cite{Koster:2016ebi} upon taking the derivatives $\frac{1}{2}\frac{\partial^2}{\partial \xi_{\ri}^a\partial \xi_{\ri}^b}$ as prescribed in \eqref{eq: specifying components by derivatives}. It will turn out to be convenient to combine the double derivative together with the prefactor \eqref{eq: contraction prefactor} into an object which we call the \textit{forming operator} $\PPP_{\phi}$ of the field $\phi$:
\beq
\PPP_{\phi}=-\frac{\la^{\alpha}\la'^{\beta}}{\langle\la\la'\rangle}\Big(-i\xi^a\frac{\partial}{\partial \theta^{\alpha a}}\Big)\Big(-i\xi^b\frac{\partial}{\partial \theta^{\beta b}}\Big)\eqndot
\eeq
Hence, with all the definitions, we can write 
\begin{equation}
 \lim_{\hexagon\rightarrow \xdot}\PPP_{\phi}\, U_{x}(\la,\la')_{\big{|}\theta=0}=h_x(\la)\textbf{W}_{\phi(x)} h_x(\la)^{-1}\eqndot
\end{equation}

\subsection{Our construction}
\label{subsec:our wilson loop construction}
We have seen how to construct a vertex for the scalar field $\phi_{ab}$ by taking two derivatives of the Wilson loop \eqref{eq: polygonal Wilson loop}. 
Similarly, we find the vertices for all other fields of $\mathcal{N}=4$ SYM by taking various combinations of derivatives of $U_{x_i}$ with respect to $x_i$ and $\theta_i$. 
We call the combination of the derivatives determining the type of field together with the contraction \eqref{eq: contraction prefactor} and the superpolarization vector \textit{the forming operator} of the specific field.
Let us now state the forming operators for the complete field content of $\calN=4$ SYM shown in \eqref{eq: alphabet of fields with polarizations}: 
\begin{equation}
 \label{eq:definitionformingfactoronshellstates}
\PPP_{D^{k_{\ri}}A_{\ri}}= -\frac{\la_{\ri}^{\alpha}\la'_{\ri}{}^{\beta}}{\abra{ \la_{\ri}}{\la'_{\ri}}} 
\big(-i\tau^\gamma_{\ri}\bar{\tau}_{\ri}^{\dot\gamma}\partial_{\ri,\gamma\dot\gamma}\big)^{k_{\ri}}
\left\{\begin{array}{ll} 
(-i\bar{\tau}_{\ri}^{\dot\alpha}\partial_{\ri,\alpha\dot\alpha})(-i\bar{\tau}_{\ri}^{\dot\beta}\partial_{\ri,\beta\dot{\beta}})& 
\text{ for } A_{\ri}=\bar{F}\\
(-i\bar{\tau}_{\ri}^{\dot\alpha}\partial_{\ri,\alpha\dot{\alpha}})(-i\xi_{\ri}^{a}\partial_{\ri,\beta a})& 
\text{ for } A_{\ri}=\bar{\psi}\\
(-i\xi_{\ri}^{a}\partial_{\ri,\alpha a})(-i\xi_{\ri}^b\partial_{\ri,\beta b})& 
\text{ for } A_{\ri}=\phi\\
(-i\xi_{\ri}^{a}\partial_{\ri,\alpha a})(-i\xi_{\ri}^{b}\partial_{\ri,\beta b})(-i\tau_{\ri}^{\gamma}\xi_{\ri}^{c}\partial_{\ri,\gamma c})&
\text{ for }A_{\ri}=\psi\\
(-i\xi_{\ri}^{a}\partial_{\ri,\alpha a})(-i\xi_{\ri}^{b}\partial_{\ri,\beta b})(-i\tau_{\ri}^{\gamma}\xi_{\ri}^{c}\partial_{\ri,\gamma c})^2& 
\text{ for } A_{\ri}=F
\end{array}\right. \eqncom
\end{equation}
where we used the abbreviations
\begin{equation}
 \partial_{\ri,\alpha\dot\alpha}\equiv \frac{\partial}{\partial x_{\ri}^{\alpha\dot\alpha}}\,,\qquad \partial_{\ri,\alpha a} \equiv \frac{\partial}{\partial \theta_{\ri}^{\alpha a}}\,.
\end{equation}

\begin{figure}[tbp]
 \centering
  \includegraphics[height=3.5cm]{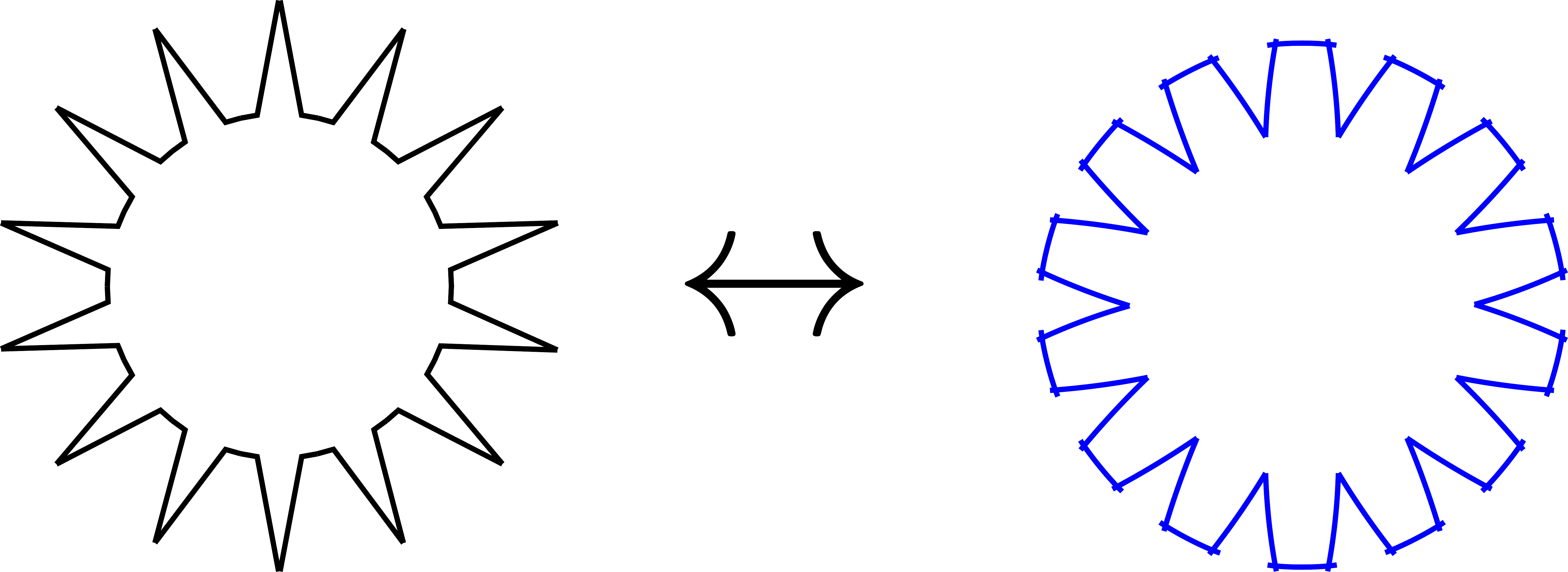}
  \caption{\it Composite operators of $L$ irreducible fields are obtained from cogwheel Wilson loops of $3L$ vertices that have $3L$ edges in twistor space.
  }
  \label{fig:CogwheelBig}
\end{figure}

In general, the geometry of the light-like Wilson loop in twistor space should allow us to take derivatives at position $x_{\ri}$ in the direction of  $\ftau_{\ri}$ and to take an operator limit where the spinor intersections $\la_{\ri}$ and $\la_{\ri}'$ coincide and are proportional to $\tau_{\ri}$. The latter condition is necessary as it must not make a difference which derivatives in \eqref{eq:definitionformingfactoronshellstates} are contracted with $\tau_{\ri}$ and which with $\la_{\ri}$ and $\la_{\ri}'$.
In appendix \ref{app:geometry}, we elaborate more on the requirements which the Wilson loop must satisfy. 
For now, suffice it to say that a Wilson loop that does the job for the composite operator of $L$ irreducible fields is a ``cogwheel'' which has $3L$ corners in position space or equivalently $3L$ edges in twistor space, see figure~\ref{fig:CogwheelBig}.%
\footnote{The number $3L$ of edges does the job for general operators, but it might not be the minimal number with this property. For special operators, less edges do in fact suffice. However, adding more edges is always possible and increases the complexity of the calculation only mildly while leaving the result for the operator vertex invariant.}
We refer also to figure~\ref{fig:CogwheelZoom} for details about the parametrization of the Wilson loop.

Finally, the operator $\mathcal{O}(x)$ is obtained by acting on the Wilson loop $\mathcal{W}$ with the forming operator $\PPP$ and then shrinking the loop to a point. As we wrote in \ref{subsec:compositeoperatorss}, we denote this operator limit by $\hexagon\rightarrow \xdot$, and hence write for the (vertex of the) composite operator:
\beq
\textbf{W}_{\mathcal{O}(x)}=\lim_{\hexagon\rightarrow \xdot}\PPP_{\mathcal{O}}\, \mathcal{W}_{\big{|}\theta=0}\,.
\eeq

The resulting vertices for each field in the composite operator are of the general form 
\begin{align}
\label{eq: minimal vertices}
 \textbf{W}_{D^{k_{\ri}}A_{\ri}(x)}&=
 \int \DD\la_{\ri} h^{-1}_x(\la_{\ri})
 \left\{\begin{array}{ll}
 \abra{\tau_{\ri}}{\la_{\ri}}^{k_{\ri}\phantom{+1}}(\bar{\tau}_{\ri}^{\dot\alpha}\partial_{i,\dot{\alpha}})^{k_{\ri}+2}\phantom{(\xi_{\ri}^a\partial_{i,a})^2}\AAA(\la_{\ri})& 
 \text{ for } A_{\ri}=\bar{F}\\
 \abra{\tau_{\ri}}{\la_{\ri}}^{k_{\ri}\phantom{+1}}(\bar{\tau}_{\ri}^{\dot\alpha}\partial_{i,\dot\alpha})^{k_{\ri}+1}(\xi_{\ri}^a\partial_{i,a})^{\phantom{1}}\AAA(\la_{\ri})& 
 \text{ for } A_{\ri}=\bar{\psi}\\
 \abra{\tau_{\ri}}{\la_{\ri}}^{k_{\ri}\phantom{+1}}(\bar{\tau}_{\ri}^{\dot\alpha}\partial_{i,\dot\alpha})^{k_{\ri}\phantom{+1}}(\xi_{\ri}^a\partial_{i,a})^2\AAA(\la_{\ri})& 
 \text{ for } A_{\ri}=\phi\\
 \abra{\tau_{\ri}}{\la_{\ri}}^{k_{\ri}+1}(\bar{\tau}_{\ri}^{\dot\alpha}\partial_{i,\dot\alpha})^{k_{\ri}\phantom{+1}}(\xi_{\ri}^a\partial_{i,a})^3\AAA(\la_{\ri})& 
 \text{ for } A_{\ri}=\psi\\
\abra{\tau_{\ri}}{\la_{\ri}}^{k_{\ri}+2}(\bar{\tau}_{\ri}^{\dot\alpha}\partial_{i,\dot\alpha})^{k_{\ri}\phantom{+1}} (\xi_{\ri}^a\partial_{i,a})^4\AAA(\la_{\ri})& 
 \text{ for } A_{\ri}=F
\end{array}
\right\}h_x(\la_{\ri})
\nonumber\\
&\phaneq+\text{terms at least quadratic in }\calA \eqncom 
\end{align}
where we used the abbreviations
\begin{equation}
\partial_{\ri,\dot\alpha}\equiv \frac{\partial}{\partial \mu_{\ri}^{\dot\alpha}}\,,\qquad
\partial_{\ri,a}\equiv \frac{\partial}{\partial \chi_{\ri}^{a}}\,.
\end{equation}
In addition to the explicitly shown terms, the vertices contain further terms which are at least quadratic in the field $\AAA$. They have partial derivatives with respect to $\mu$ and $\chi$ act on up to $k_{\ri}+2$, $k_{\ri}+2$, $k_{\ri}+2$, $k_{\ri}+3$ and $k_{\ri}+4$ different $\AAA$, respectively.
The higher order terms are straightforward to obtain using the product rule.
The previously treated example of the scalar field was already given in \eqref{eq: vertex scalar}.
We give further explicit examples of vertices for fields in appendix \ref{app:formfactordatamine}. 

The vertex for an operator $\mathcal{O}(x)=\Tr(D^{k_1}A_1(x)\cdots D^{k_L}A_L(x))$ is then given by 
\begin{equation}
 \label{eq: operator vertex}
 \textbf{W}_{\mathcal{O}(x)}=\Tr(\textbf{W}_{D^{k_1}A_1(x)}\cdots\textbf{W}_{D^{k_L}A_L(x)})\eqncom
\end{equation}
where the frames $h_x$ and inverse frames $h_x^{-1}$ in \eqref{eq: minimal vertices} combine to parallel propagators $U_x$ according to \eqref{eq: parallel propagator in terms of frames}.

Note that we can also refrain from setting $\theta=0$ to obtain the vertex for the chiral part of a supermultiplet of which $\mathcal{O}$ is the lowest component.

\section{MHV form factors}
\label{sec:MHVformfactors}

Having constructed the vertices for all composite operators, we can use them to derive their tree-level MHV form factors.
This also serves as a cross check of our construction.

\subsection{All minimal form factors}
\label{subsec:minimalformfactors}

As a warm-up, we rederive the minimal form factor of a generic operator from the field vertices \eqref{eq: minimal vertices}.
Taking the product of the respective terms, we find that the frames $h_x$ and inverse frames $h_x^{-1}$ combine into parallel propagators $U_x$. At lowest valency, the parallel propagator is simply $1$ and we only require the terms explicitly shown in \eqref{eq: minimal vertices}, so that
\begin{multline}
\label{eq: vertex for minimal valency}
 \textbf{W}_{\Tr(D^{k_1}A_1\cdots D^{k_L}A_L)(x)}\Big|_{L\text{-valent}}\\
 =\prod_{\ri=1}^L \int \DD\la_{\ri}  \left\{\begin{array}{ll}
 \abra{\tau_{\ri}}{\la_{\ri}}^{k_{\ri}\phantom{+1}}(\bar{\tau}_{\ri}^{\dot\alpha}\partial_{\ri,\dot{\alpha}})^{k_{\ri}+2}& 
 \, \text{ for }A_{\ri}=\bar{F}\\
 \abra{\tau_{\ri}}{\la_{\ri}}^{k_{\ri}\phantom{+1}}(\bar{\tau}_{\ri}^{\dot\alpha}\partial_{\ri,\dot\alpha})^{k_{\ri}+1}(\xi_{\ri}^a\partial_{\ri,a})& 
 \, \text{ for }A_{\ri}=\bar{\psi}\\
 \abra{\tau_{\ri}}{\la_{\ri}}^{k_{\ri}\phantom{+1}}(\bar{\tau}_{\ri}^{\dot\alpha}\partial_{\ri,\dot\alpha})^{k_{\ri}\phantom{+1}}(\xi_{\ri}^a\partial_{\ri,a})^2& 
 \, \text{ for }A_{\ri}=\phi\\
 \abra{\tau_{\ri}}{\la_{\ri}}^{k_{\ri}+1}(\bar{\tau}_{\ri}^{\dot\alpha}\partial_{\ri,\dot\alpha})^{k_{\ri}\phantom{+1}}(\xi_{\ri}^a\partial_{\ri,a})^3& 
 \, \text{ for }A_{\ri}=\psi\\
\abra{\tau_{\ri}}{\la_{\ri}}^{k_{\ri}+2}(\bar{\tau}_{\ri}^{\dot\alpha}\partial_{\ri,\dot\alpha})^{k_{\ri}\phantom{+1}} (\xi_{\ri}^a\partial_{\ri,a})^4& 
 \, \text{ for }A_{\ri}=F
\end{array}
\right\}\AAA(\la_{\ri})_{\big{|}\theta=0} \eqncom
\end{multline}
where the product is understood to be ordered and $\AAA(\la_{\ri})\equiv \AAA(\calZ_{x}(\la_{\ri}))$.

From the operator vertices \eqref{eq: vertex for minimal valency}, we can straightforwardly derive the minimal tree-level position-space form factors $\calF_{\calO}(1,\ldots, L;x)$ by replacing the $\AAA(\la_\ri)$  with the on-shell states $\AAA_{\fP_\rj}(\la_\ri)$ \eqref{eq:definitiononshellmomentumeigenstates}. There are $L$ distinct but cyclically related ways to planarly connect the $L$ $\AAA$'s in \eqref{eq: vertex for minimal valency} to external (super)momentum eigenstates \eqref{eq:definitiononshellmomentumeigenstates}. They are entirely determined by choosing to connect the field $1$ to the external field $\rj+1$, $2$ to $\rj+2$ and so on. The integration over the spinors completely factorizes. As an example, consider the case $\bar F$ or helicity one: the contributing factor is (dropping the superfluous indices)
\begin{multline}
\label{eq:deltaintegration}
 \int \DD\la \abra{\tau}{\la}^k(\bar{\tau}^{\dot\alpha}\partial_{\dot \alpha})^{k+2}2\pi i\int_{\mathbb{C}} \frac{\dd s}{s}\e^{s(\mu^{\dot\alpha}\bar{p}_{\dot\alpha}+\chi^a\eta_a)}\bar\delta^2(s\lambda-p)_{\big{|}\theta=0}\\
 =\int_{\mathbb{C}^2} \dd u_1\dd u_2 \abra{\tau}{u}^k\sbra{\bar p}{\bar\tau}^{k+2}\e^{ix^{\alpha\dot\alpha}u_\alpha \bar{p}_{\dot\alpha}}\bar\delta^2(u-p)=\abra{\tau}{p}^k\sbra{\bar p}{\bar \tau}^{k+2}\e^{ix\fp}\,,
\end{multline}
where we parametrized $\la=(1,u_2)$, renamed $s\rightarrow u_1$ and rescaled $u_2\rightarrow u_1u_2^{-1}$.
A similar calculation for the other cases shows that the insertion of \eqref{eq:definitiononshellmomentumeigenstates} and the subsequent integration over $s_\ri$ and $\la_\ri$ effectively replaces
\begin{equation}
 \lambda_{\ri,\alpha}\longrightarrow p_{\ri+\rj,\alpha} \eqncom \qquad 
 \partial_{\ri,\dot\alpha} \longrightarrow \bar{p}_{\ri+\rj,\dot\alpha}\eqncom \qquad 
\partial_{\ri,a}\longrightarrow \eta_{\ri+\rj,a}
\end{equation}
 in \eqref{eq: vertex for minimal valency}.
 
Finally, the minimal tree-level form factor in momentum space is obtained after Fourier transforming: 
\begin{equation}
 \label{eq: minimal form factor in momentum space}
\begin{multlined}
\calF_{\Tr(D^{k_1}A_1\cdots D^{k_L}A_L)}(1,\dots,L;\fq)=\int \frac{\dd^4 x}{(2\pi)^4}\e^{-ix\fq} \calF_{\Tr(D^{k_1}A_1\dots D^{k_L}A_L)}(1,\dots,L;x)\\
=\sum_{j=0}^{L-1}\prod_{\ri=1}^L   \left\{\begin{array}{ll}
 {\abra{\tau_{\ri}}{p_{\ri+\rj}}}^{k_{\ri}\phantom{+1}}{\sbra{\bar{p}_{\ri+\rj}}{\bar{\tau}_{\ri}}}^{k_{\ri}+2} & \text{ for } A_{\ri}=\bar{F}\\
 \abra{\tau_{\ri}}{p_{\ri+\rj}}^{k_{\ri}\phantom{+1}}\sbra{\bar{p}_{\ri+\rj}}{\bar{\tau}_{\ri}}^{k_{\ri}+1}\{\xi_{\ri}\eta_{\ri+\rj}\}& \text{ for }A_{\ri}=\bar{\psi}\\
 \abra{\tau_{\ri}}{p_{\ri+\rj}}^{k_{\ri}\phantom{+1}}\sbra{\bar{p}_{\ri+\rj}}{\bar{\tau}_{\ri}}^{k_{\ri}\phantom{+1}}\{\xi_{\ri}\eta_{\ri+\rj}\}^2& \text{ for } A_{\ri}=\phi\\
 \abra{\tau_{\ri}}{p_{\ri+\rj}}^{k_{\ri}+1}\sbra{\bar{p}_{\ri+\rj}}{\bar{\tau}_{\ri}}^{k_{\ri}\phantom{+1}}\{\xi_{\ri}\eta_{\ri+\rj}\}^3& \text{ for } A_{\ri}=\psi\\
 \abra{\tau_{\ri}}{p_{\ri+\rj}}^{k_{\ri}+2}\sbra{\bar{p}_{\ri+\rj}}{\bar{\tau}_{\ri}}^{k_{\ri}\phantom{+1}}\{\xi_{\ri}\eta_{\ri+\rj}\}^4& \text{ for } A_{\ri}=F
\end{array}
\right\} \delta^4\left(\fq-\sum_{\ri=1}^L \fp_{\ri}\right)\eqndot
\end{multlined}
\end{equation}
This result perfectly agrees with the result originally obtained in \cite{Wilhelm:2014qua}.

For the minimal form factor of the operator $\calO'=\frac{1}{2}\Tr(\phi_{ab}^2)$ in \cite{Koster:2016ebi}, we find by using the change of basis \eqref{eq: specifying components by derivatives} from \eqref{eq: minimal form factor in momentum space}:
\beq
\begin{split}
\calF_{\calO'}(1,2;\fq)&=\frac{1}{2}\frac{\partial^2}{\partial \xi_1^a\partial \xi_1^b}\frac{1}{2}\frac{\partial^2}{\partial \xi_2^a\partial \xi_2^b}\frac{1}{2}\left(\{\xi_1\eta_1\}^2\{\xi_2\eta_2\}^2+\{\xi_1\eta_2\}^2\{\xi_2\eta_1\}^2\right)\delta^4\left(\fq-\fp_1-\fp_2\right)\\
&=\eta_{1a}\eta_{1b} \eta_{2a}\eta_{2b}\delta^4\left(\fq-\fp_1-\fp_2\right)
\,.
\end{split}
\eeq
Hence, the only possible outgoing states in this case are two scalars $1^{\phi_{ab}}$ and $2^{\phi_{ab}}$ and we obtain the desired result.

\subsection{All MHV form factors}
\label{subsec: all MHV form factors}

Using the vertices constructed in section \ref{sec:construction}, we can also derive the general tree-level $n$-point MHV form factors of the  composite operators.
We treated the example of the $n$-point MHV form factor of the operator $\calO'=\frac{1}{2}\Tr(\phi_{ab}^2)$ in \cite{Koster:2016ebi}.
Let us now calculate the $n$-point MHV form factor for a generic single-trace operator containing the fields in \eqref{eq: alphabet of fields with polarizations}. 
We first write down the result and then present its derivation.

The $n$-point tree-level MHV form factor of a generic single-trace operator  $\calO$ \eqref{eq:composite operator} in $\calN=4$ SYM reads
\beq
\begin{split}
\label{eq: general MHV form factor}
\calF_{\mathcal{O}}(1,\dots,n;\fq)
 &=\frac{\delta^4(\fq-\sum_{i=1}^n \fp_i)}{\prod_{i=1}^n\abra{i}{(i+1)}} \sum_{\{\notA_{a,b}\}}
 \prod_{j=1}^L 
 \abra{p_{\notA_{j,N_j}}}{p_{\notA_{j+1,1}}}
\left(\prod_{k=2}^{N_j-1}\abra{\tau_j}{p_{\notA_{j,k}}}\right)\\
 &\phaneq
 \sum_{\sigma\in S_{N_j}}
 \frac{\{\xi_j \eta_{\sigma(\notA_{j,1})}\}\cdots\{\xi_j \eta_{\sigma(\notA_{j,n_{\theta_j}})}\}\sbra{\bar{p}_{\sigma(\notA_{j,n_{\theta_j}+1})}}{\bar\tau_j}\cdots\sbra{\bar{p}_{\sigma(\notA_{j,N_j})}}{\bar\tau_j}}{M(\{\notA_{j,1},\dots,\notA_{j,N_j}\})!}
\\
 &\phaneq
 +\text{cyclic permutations}
 \eqndot
\end{split}
\eeq
We are using the following notation:
\begin{enumerate}
\item  We denote by $N_i$ the total number of indices $\dot\alpha$ and $a$ of the field $D^{k_i}A_i$ in \eqref{eq: alphabet of fields with polarizations}, which is equal to the number of derivatives required to generate the field $D^{k_i}A_i$ in \eqref{eq:definitionformingfactoronshellstates}. Specifically, for $A_i=\bar F, \bar \psi $ or $\bar \phi$, we have $N_i=k_i+2$. Otherwise, $N_i=k_i+3$ for $A=\psi$ and $N_i=k_i+4$ for $A_i=F$. 
Moreover, we denote by $n_{\theta_\ri}$ the number of indices $a$ of the field $D^{k_i}A_i$ in \eqref{eq: alphabet of fields with polarizations}, which is equal to the number of $\frac{\partial}{\partial\theta_\ri}$ derivatives required to generate the field $D^{k_i}A_i$ in \eqref{eq:definitionformingfactoronshellstates}. Concretely, $n_{\theta_\ri}=0,1,2,3,4$ for $A_i=\bar F, \bar \psi, \phi, \psi, F $, respectively.
\item The sum in \eqref{eq: general MHV form factor} is over all sets $\{B_{i,j}|i=1,\ldots,L, j=1,\ldots, N_i\}$ with $1\leq \notA_{1,1}\leq \dots \leq \notA_{1,N_1}<\dots<\notA_{L,1}\leq \dots \leq \notA_{L,N_L}\leq n$. 
\item 
We denote by $M(\{\notA_{i,1},\notA_{i,2}, \dots , \notA_{i,N_i}\})$ the set of multiplicities of the entries in the original set. For example, $M(\{2,3,3,7,9,9,9\})=\{1,2,1,3\}$. 
We define $\{a,b,c,\dots\}!= a!b!c!\cdots$, for example $ M(\{2,3,3,7,9,9,9\})!=1!2!1!3!=12$.
Note that for $n_{\theta_\ri}=0$ or $n_{\theta_\ri}=N_\ri$ the sum over all permutations reduces to $N_{\ri}!$ and the total numeric prefactor in the second line of \eqref{eq: general MHV form factor} becomes a multinomial coefficient.
\end{enumerate}

Before we discuss the proof of \eqref{eq: general MHV form factor}, let us make some remarks. To start, we emphasize that our result \eqref{eq: general MHV form factor} is the first complete computation of the MHV tree-level form factors of \textit{all} $\calN=4$ SYM composite operators. 
In particular, \eqref{eq: general MHV form factor} is consistent with all available computations for specific operators -- it agrees with the results of \cite{Brandhuber:2011tv} for the operators in the stress-tensor supermultiplet, with those of \cite{Engelund:2012re} for operators in the $SU(2)$ sector and those of \cite{Engelund:2012re} for twist-two operators in the $SL(2)$ sector. 
Let see how \eqref{eq: general MHV form factor} reduces to the result for scalars in some more detail.
For scalar fields, the second line in \eqref{eq: general MHV form factor} reduces to 
\begin{equation}
\label{eq: X factor scalar}
\begin{aligned}
 \{\xi_i \eta_{\notA_{i,1}}\} \{\xi_i \eta_{\notA_{i,2}}\} (\delta_{\notA_{i,1}=\notA_{i,2}} + 2\delta_{\notA_{i,1}\neq \notA_{i,2}})\eqndot 
 \end{aligned}
\end{equation}
Upon taking derivatives with respect to $\xi_i^a$ as specified in \eqref{eq: specifying components by derivatives}, one is essentially left with the expression found in \cite{Engelund:2012re}.
Moreover, we have checked a wide range of cases which are not available in the literature using Feynman diagrams.

We can derive \eqref{eq: general MHV form factor} in two different ways.
The first way closely follows our derivation of the MHV form factor of $\calO'=\frac{1}{2}\Tr(\phi_{ab}\phi_{ab})$ in \cite{Koster:2016ebi}.
For every given field $A_i$ in \eqref{eq:definitionformingfactoronshellstates}, we can apply the derivatives to the Wilson loop and perform the operator limit to obtain the corresponding vertices, as was done for several examples in appendix \ref{app:formfactordatamine}. 
We can then insert external momentum eigenstates \eqref{eq:definitiononshellmomentumeigenstates} and perform the Fourier transformation to arrive at \eqref{eq: general MHV form factor}.%
\footnote{Heuristically, the MHV denominator in \eqref{eq: general MHV form factor} stems from the combined parallel propagators $U$. 
The second term stems from the numerators of the $U$'s between the different irreducible fields $D^{k_i}A_i$.
The third term stems from the prefactors of the derivatives and the second line accounts for the combinatorics of acting with the derivatives.}
The second way to derive \eqref{eq: general MHV form factor} is to insert momentum eigenstates into the Wilson loop vertex to compute the form factor of the Wilson loop, then act with the derivatives and perform the operator limit in the end. The second derivation is given in full detail in appendix \ref{app: derivation}. 

Finally, let us remark that one could also obtain form factors of the chiral parts of the supermultiplets that contain the operator $\mathcal{O}$ as lowest component by taking a suitable fermionic Fourier transformation with respect to $\theta$ instead of setting it to zero in analogy to what was done for the stress-tensor supermultiplet in \cite{Brandhuber:2011tv}.

\section{Conclusion and outlook}\label{sec:conclusion}
In this paper, we have derived expressions for all gauge-invariant local composite operators of $\calN=4$ SYM in twistor space and obtained all their tree-level MHV super form factors, thus extending the formalism developed in \cite{Koster:2016ebi}.
In section \ref{sec:construction}, we have constructed the twistor-space vertices of the composite operators from the cogwheel light-like Wilson loops presented in appendix \ref{app:geometry}. In section \ref{sec:MHVformfactors} and appendix \ref{app: derivation}, we have shown that they immediately generate all tree-level MHV form factors by simply inserting on-shell external particles. In addition to providing an important new result, namely all the tree-level MHV form factors of $\calN=4$ SYM shown in \eqref{eq: general MHV form factor}, our calculations reveal a striking analogy between form factors and amplitudes in twistor space.

In a forthcoming paper \cite{Paper3}, we build further on the foundations laid in this paper by extending the framework to tree-level N$^k$MHV form factors. Starting from the cogwheel Wilson loops, we use an inverse soft limit for the operator vertices which allows us to compute form factors beyond MHV level in a similar fashion as amplitudes. In addition, we will demonstrate that the formalism can be applied to correlation functions and in particular to straightforwardly show the cancellation that was mentioned in  \cite{Koster:2014fva}.

There exists an interesting alternative to the twistor space formalism, the so-called Lorentz harmonic chiral (LHC) superspace put forward in \cite{Chicherin:2016fac, Chicherin:2016fbj}.  It is,  according to the authors, conceptually simpler. Our results on composite operators reported in \cite{Koster:2016ebi} have recently been confirmed in \cite{Chicherin:2016soh} using the LHC. It would be very interesting to also put the expanded results on composite operators in the current paper to a test. Even more exciting would be a re-derivation of equation \eqref{eq: general MHV form factor} in this formalism.

The methods used in this paper lend themselves to several possible further extensions. A first direction of future research would be to return to the original motivation of \cite{Koster:2014fva} and to study the integrability of $\calN=4$ SYM directly from twistor space. In particular, (partial) Yangian symmetry and integrability have already been uncovered in tree-level form factors \cite{Frassek:2015rka} and it would be interesting to see whether twistor-space techniques can be used to push that development further. 
In the intermediate steps of our derivation of all MHV form factors, we have moreover calculated the form factors of our decorated Wilson loops.
Special kinematic limits of form factors of certain decorated Wilson loops in space-time, namely light-ray operators, were previously studied in some subsectors in \cite{Derkachov:2013bda}, where they were related to the dual conformal invariance of the dilatation operator and to an integrability-based construction of the one-loop eigenstates under renormalization.
It would be interesting to see this relation also in twistor space.
In a related line of study, one should investigate multiloop computations purely in twistor space. For this, one would need to understand the line-splitting of \cite{Koster:2014fva} better, see also \cite{Adamo:2013cra} for a discussion on this point. Note that our Wilson loop already implements the line-splitting of \cite{Koster:2014fva}. 
Finally, one could try to apply position twistor-space techniques to theories other than $\calN=4$ SYM. For scattering amplitudes, this was already done with momentum twistors starting in \cite{Hodges:2009hk}. In \cite{Boels:2006ir}, the problem of writing down the action of arbitrary $\calN=1$ and $\calN=2$ SYM in 4d was in principle solved. It should therefore be possible to apply these results to compute form factors using position-space twistors in those theories as well. 

\section*{Acknowledgments}
\label{sec:acknowledgments}
We are thankful to Tim Adamo, Simon Caron-Huot and especially Lionel Mason for insightful comments and  discussions. This research is supported in part by the SFB 647 \emph{``Raum-Zeit-Materie. Analytische und Geometrische Strukturen''}. V.M.\ is also supported by the PRISMA cluster of excellence at the Johannes Gutenberg University in Mainz.
V.M.\ would like to thank the Simons Summer Workshop 2015, where part of this work was performed. 
M.W.\ was supported  in  part  by  DFF-FNU  through grant number DFF-4002-00037. L.K.\ would like to thank the CRST at Queen Mary University of London and especially Gabriele Travaglini for hospitality during a crucial stage of preparation of this work.

\appendix
\addtocontents{toc}{\protect\setcounter{tocdepth}{1}}

\section{Geometry of the Wilson loop}
\label{app:geometry}

In this appendix, we present explicitly the  geometry of the Wilson loop that is used in the construction of the composite operators of section \ref{sec:construction}. Specifically, we start with a list of requirements that the Wilson loop needs to satisfy, then show our solution and finish with the operator limit -- the procedure that sends the Wilson loop to a point.

As explained in subsection \ref{subsec:compositeoperatorss}, we need to make sure that the local operators do not depend on the edges of the Wilson loop $\la_i$ or $\la_i'$ after the loop has been shrunk. 
Moreover, it must not matter which derivatives in \eqref{eq:definitionformingfactoronshellstates} are contracted with $\la_i$, which with $\la'_i$ and which with the polarization vectors $\tau_i$ from \eqref{eq: alphabet of fields with polarizations}.
The solution is to shrink the Wilson loop in such a way as to achieve $\la_i\rightarrow \tau_i$ and $\la_i'\rightarrow \tau_i$. This needs to be done for each edge of the Wilson loop that is acted upon by derivatives. We thus see that we need to add extra edges to the Wilson loop that will not be acted upon by derivatives, i.e.\ that will not carry any irreducible fields \eqref{eq: alphabet of fields with polarizations}. Were it not so, i.e.\ if we acted with derivatives on each edge of the Wilson loop, then in the operator limit, all the corners of the loop would have to be identical. This would then imply that we have just one independent supertwistor on the loop in that limit, see figure \ref{fig:operatorlimit}, but we need to have at least two in order for the line $x$ to be well defined.
Finally, we also need to be able to take derivatives of $x_i$ in the direction $\tau_i$, i.e.\ to infinitesimally vary $x_i$ in this direction without destroying the light-like nature of the Wilson loop. This constraints the relative positions of the points neighboring $x_i$.

\begin{figure}[htbp]
 \centering
  \includegraphics[height=3.7cm]{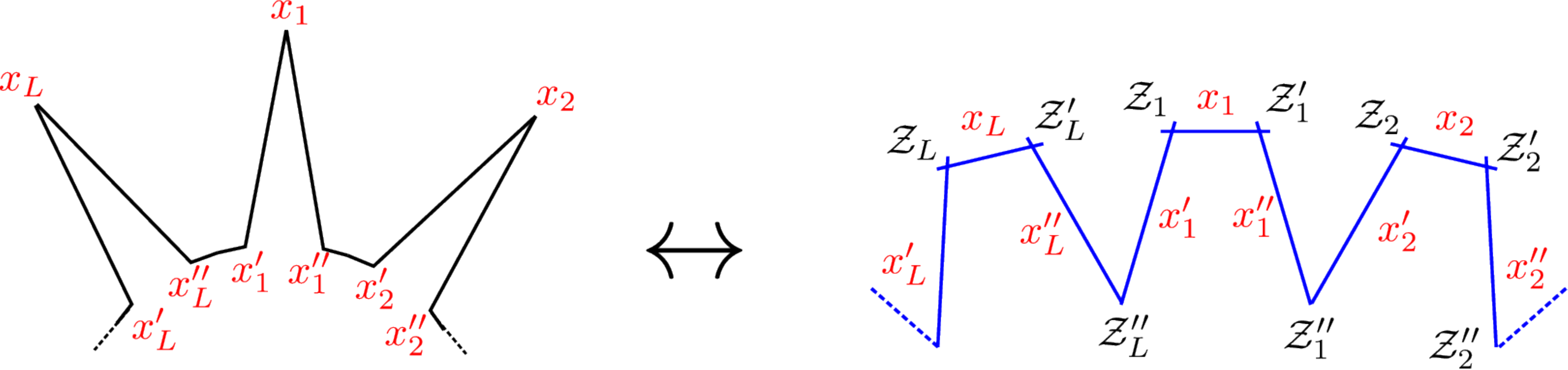}
  \caption{\it The geometry of the light-like Wilson loop. }
  \label{fig:CogwheelZoom}
\end{figure}

The simplest Wilson loop geometry that we found involves $3L$ edges for a general composite operator $\calO$ of length $L$ \eqref{eq:composite operator}. The shape of our Wilson loop is reminiscent of a cogwheel. Specifically, we consider a light-like Wilson loop as shown in figure~\ref{fig:CogwheelBig} with $3L$ points, or corners, in space-time. These points are labeled as $x_i$, $x'_i$ and $x''_i$ with $i=1,\ldots,L$, see figure \ref{fig:CogwheelZoom}, and they are ordered as $x'_1,x_1,x''_1,\ldots, x'_L,x_L,x''_L$. 
The (super) light-like condition implies that any two neighboring points $(x,\theta)$ and $(y, \vartheta)$ must satisfy $(x-y)^2=0$ and $(x-y)^{\alpha\dot\alpha}(\theta-\vartheta)_{\alpha}^{\phantom{\alpha}a}=0$. 
For the cogwheel Wilson loop, we solve these constraints as follows. We let $(x,\theta)$ be the center of the loop and parametrize
\begin{equation}
\label{eq:Cogwheelparametrization}
\begin{aligned}
&x_i^{\alpha\dot{\alpha}}=x^{\alpha\dot{\alpha}}+\m_i^{\alpha}\bm_i^{\dot{\alpha}}\eqncom&\quad &{x_i'}^{\alpha\dot{\alpha}}=x^{\alpha\dot{\alpha}}+\n_i^{\alpha}\bm_i^{\dot{\alpha}}\eqncom&\quad &{x_i''}^{\alpha\dot{\alpha}}=x^{\alpha\dot{\alpha}}+\n_{i+1}^{\alpha}\bm_i^{\dot{\alpha}}\eqncom&
\\
&\theta_i^{\alpha a }= \theta + \m_i^{\alpha} \xi_i^a\eqncom&  &{\theta_i'}^{\alpha a }=\theta + \n_i^{\alpha} \xi_i^a\eqncom&  &{\theta_i''}^{\alpha a }= \theta + \n_{i+1}^{\alpha} \xi_i^a\eqncom&
\end{aligned}
\end{equation}
where $\m_i$, $\bm_i$ and $\n_i$ are complex spinors and $\xi_i$ are Gra\ss mann parameters.
Thus, \eqref{eq:Cogwheelparametrization} ensures that the loop is light-like. We express the differences between the points as
\begin{equation}
\begin{split}
&\lambda_i\bar{\lambda}_i=x'_{i}-x_i=(\n_i-\m_i)\bm_i\eqncom\\
&\lambda_i'\bar{\lambda}_i'=x_{i}-x''_i=(\m_i-\n_{i+1})\bm_i\eqncom\\
&\lambda_i''\bar{\lambda}_i''=x''_{i}-x'_{i+1}=\n_{i+1}(\bm_i-\bm_{i+1})\eqndot
\end{split}
\end{equation}
We can choose, up to rescaling, to satisfy the above equations via
\begin{align}
\label{eq:definitionoflamdalambdaprimeandsecond}
&\lambda_i=\n_i-\m_i\eqncom& &\lambda_i'=\m_i-\n_{i+1}\eqncom& &\lambda_i''=\n_{i+1}\eqncom&\nonumber\\ &\bar{\lambda}_i=\bm_i\eqncom&
 &\bar{\lambda}_i'=\bm_i\eqncom&
 &\bar{\lambda}_i''=\bm_i-\bm_{i+1}\eqndot&
\end{align}
The twistors that correspond to the intersection of the lines are then
\begin{equation}
\label{eq:intersectiontwistorsdefinition}
\begin{split}
x'_i\cap x_i&=\calZ_i=(\lambda_i, i(x+\m_i\bm_i) \lambda_i, i(\theta+\m_i\xi_i)\lambda_i)\eqncom\\
x_i\cap x''_i&=\calZ_i'=(\lambda_i', i(x+\m_i\bm_i) \lambda_i',i(\theta+\m_i\xi_i)\lambda_i')\eqncom\\
x''_i\cap x'_{i+1}&=\calZ_i''=(\lambda_i'', i(x+\n_{i+1}\bm_i) \lambda_i'',i(\theta+\n_{i+1}\xi_i)\lambda_k'')\,,
\end{split}
\end{equation}
where the index contractions are left implicit. Putting everything together, we write down our cogwheel Wilson loops as
\begin{equation}
\label{eq:finaldefinitionWilsonloop}
\begin{split}
 \WWW(x_1',x_1,x_1'',\ldots, x_L',x_L,x_L'')= \Tr\Big[&U_{x_1'}(\calZ_L'',\calZ_1) U_{x_1}(\calZ_1,\calZ_1')U_{x_1''}(\calZ_1',\calZ_1'')\\&  U_{x_2'}(\calZ_1',\calZ_2)U_{x_2}(\calZ_1,\calZ_1')U_{x_2''}(\calZ_1',\calZ_2'') \cdots \\&\cdots U_{x_L'}(\calZ_{L-1}'',\calZ_L) U_{x_L}(\calZ_{L},\calZ_L') U_{x_L''}(\calZ_{L}',\calZ_L'') \Big]\eqncom
\end{split}
\end{equation}
cf.\ figure \ref{fig:CogwheelZoom}. We act on this Wilson loop with derivatives, as in \eqref{derivativeonUU}, but only on the unprimed edges.

\begin{figure}[htbp]
 \centering
  \includegraphics[height=3cm]{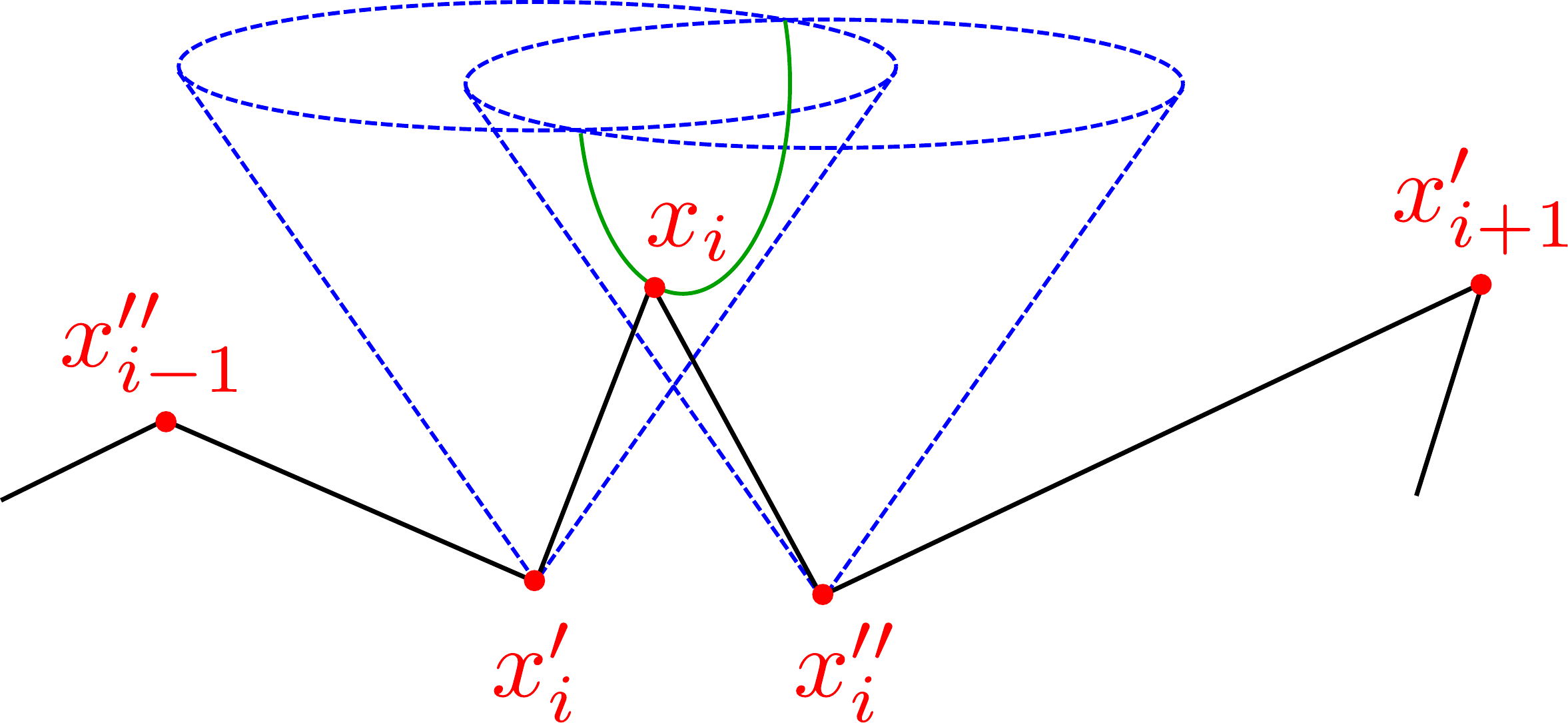}
  \caption{\it The variation of the Wilson loop has to preserve its light-like structure, thus constraining the point $x_i$ to lie on the intersection of the two light cones emanating from $x_{i-1}$ and $x_{i+1}$, depicted here in green.}
  \label{fig:LightLikeDeformation}
\end{figure}
In order to implement operators with covariant derivatives, we need to consider infinitesimal variations of the loop, specifically of the points $x_i$. To preserve the light-like nature of the loop at first order, see figure~\ref{fig:LightLikeDeformation}, the variations need to be of the type
\begin{equation}
\label{eq:definitionofthepolarizationvectors}
\delta x_i=a_i \lambda_i\bar{\lambda}_i'+b_i\lambda_i'\bar{\lambda}_i=\big(a_i(\n_i-\m_i)+b_i(\m_i-\n_{i+1})\big)\bm_i\equiv c_i \tau_i\bar{\tau}_i\eqncom
\end{equation}
where $a_i$, $b_i$ and $c_i$ are arbitrary infinitesimal parameters. The spinors $\tau_i$ and $\bar{\tau}_i$ are identified as the polarization vectors \eqref{eq: alphabet of fields with polarizations}. Equation \eqref{eq:definitionofthepolarizationvectors} can be used to solve for $\m_i$, $\n_i$ and $\bm_i$ as functions of $\tau_i$ and $\bar{\tau}_i$. There are clearly many solutions, but as we are only interested in them in the limit in which the Wilson loop shrinks to a point, we shall abstain from presenting them here.

We now want to discuss how the loop is to be shrunk to a point, i.e.\ \textit{the operator limit}. With that goal in mind, we first rescale the spinors $\m_i$, $\bm_i$ and $\n_i$ by $u$. For the intersection twistors $\calZ_i=(\lambda_i,\mu_i,\chi_i)$ of \eqref{eq:intersectiontwistorsdefinition}, this has the effect
\begin{equation}
\lambda_i\rightarrow u \lambda_i\eqncom \qquad \mu_i\rightarrow i(x+u^2\m_i\bm_i)u\lambda_i\eqncom \qquad \chi_i\rightarrow i(\theta+u^2\m_i\xi_i)u\lambda_i\eqncom
\end{equation}
and similarly for the $\calZ_i'$ and $\calZ_i''$.
Since the twistors  are projective quantities, the overall $u$ is irrelevant. Hence, rescaling the $\calZ_i$ leaves the $\lambda_i$ invariant, and in the limit $u\rightarrow 0$ we get
\begin{equation}
\label{eq:oplimit1}
\calZ_i\rightarrow (\lambda_i, ix\lambda_i, i\theta \lambda_i)\eqncom \qquad \calZ_i'\rightarrow (\lambda_i', ix\lambda_i',i \theta\lambda_i')\eqncom\qquad \calZ_i''\rightarrow (\lambda_i'', ix\lambda_i'', i\theta \lambda_i'')\eqncom
\end{equation}
i.e.\ all the intersection twistors lie on the same line.
\begin{figure}[htbp]
 \centering
  \includegraphics[height=4.2cm]{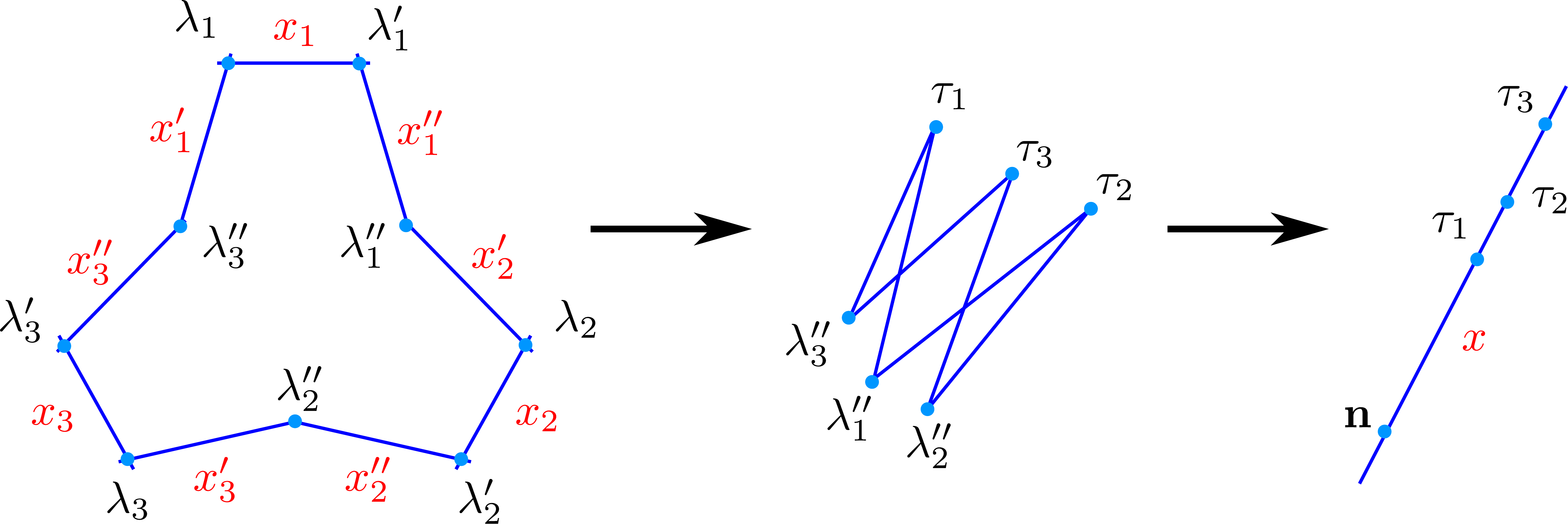}
  \caption{\it This figure illustrates the operator limit for $L=3$. To visualize the process a bit better, as a first step, we set $\lambda_i=\la_i'=\tau_i$ while bringing the $x_i$ closer to each other. The second step then just sends all $x_i$ to $x$ and $\lambda_{i}''\rightarrow \n$. 
  }
  \label{fig:operatorlimitapp}
\end{figure}
The above limit does not yet realize $\lambda_i\parallel\lambda_i'\parallel\tau_i$.
This can be achieved by setting all spinors $\n_i=\n$ to be equal. 
Due to \eqref{eq:definitionoflamdalambdaprimeandsecond} and \eqref{eq:definitionofthepolarizationvectors}, this sets $\lambda_i$, $\lambda_i'$ and  $\tau_i$ equal up to rescaling and we choose them to be equal. Summarizing, we have in the operator limit:
\begin{equation}
\label{eq:oplimit2}
\lambda_i\rightarrow \tau_i\,,\qquad  \lambda_i' \rightarrow \tau_i\,,\qquad \lambda_i''\rightarrow \n \eqndot
\end{equation}
In addition, due to \eqref{eq:definitionofthepolarizationvectors}, in the operator limit we have
\begin{equation}
\tau_i=\m_i-\n\eqncom\qquad \bar{\tau}_i=\bm_i\eqndot
\end{equation}
We illustrate the operator limit geometrically in figure~\ref{fig:operatorlimitapp}.

\section{Operator vertices}
\label{app:formfactordatamine}

In this appendix, for the convenience of the reader, we have worked out the results of applying the derivative operators \eqref{eq:definitionformingfactoronshellstates} and the operator limit \eqref{eq:oplimit1}, \eqref{eq:oplimit2} to obtain the field vertices $\mathbf{W}_{D^{k_i}A_i(x)}$ in several explicit examples.
Multiplying them together according to \eqref{eq: operator vertex} leads to the corresponding operator vertices.

For an anti-fermion, written as $\bar{\psi}=\bar{\tau}^{\dot\alpha}\xi^a\bar{\psi}_{a\dot\alpha}(x)$, the field vertex reads 
\begin{equation}\label{eq: anitfermion}
\begin{aligned}
 \mathbf{W}_{\bar{\psi}}=&\int \DD\la h_x^{-1}(\la) \bar{\tau}^{\dot\alpha}\xi^a\frac{\partial^2 \AAA(\la)}{\partial \chi^a \partial\mu^{\dot\alpha}}h_x(\la)
 +\int \DD\la\DD\la' h_x^{-1}(\la) \xi^a\frac{\partial \AAA(\la)}{\partial\chi^a}\frac{U_x(\la,\la')}{\abra{\la}{\la'}}\bar{\tau}^{\dot\alpha}\frac{\partial \AAA(\la')}{\partial\mu'^{\dot\alpha}}h_x(\la')\\
 &+\int \DD\la\DD\la' h_x^{-1}(\la) \bar{\tau}^{\dot\alpha}\frac{\partial \AAA(\la)}{\partial\mu^{\dot\alpha}}\frac{U_x(\la,\la')}{\abra{\la}{\la'}}\xi^a\frac{\partial \AAA(\la')}{\partial\chi'^{a}}h_x(\la')
 \eqndot
\end{aligned}
\end{equation}
The anti-self-dual part of the field strength $\bar{F}=\bar{\tau}^{\dot\alpha}\bar{\tau}^{\dot\beta}\bar{F}_{\dot\alpha\dot\beta}(x)$ has the vertex
\begin{equation}\label{eq: asdfieldstrength}
\begin{aligned}
 \mathbf{W}_{\bar{F}}= &\int \DD\la h^{-1}_x(\la) \bar{\tau}^{\dot\alpha}\bar{\tau}^{\dot\beta}\frac{\partial^2 \AAA(\la)}{\partial\mu^{\dot\alpha} \partial\mu^{\dot\beta}}h_x(\la)
 \\
  &
  +2\int \DD\la\DD\la' h^{-1}_x(\la) \bar{\tau}^{\dot\alpha}\frac{\partial \AAA(\la)}{\partial\mu^{\dot\alpha}}\frac{U_x(\la,\la')}{\abra{\la}{\la'}}\bar{\tau}^{\dot\beta}\frac{\partial \AAA(\la')}{\partial\mu'^{\dot\beta}}h_x(\la')\eqndot
\end{aligned}
\end{equation}
The vertex of a fermion $\psi=-\tau^\alpha\xi^a\xi^b\xi^c\psi_{abc\alpha}(x)$ is
\begin{equation}\label{eq: fermion}
\begin{aligned}
& \mathbf{W}_{\psi}=-\int \DD\la h^{-1}_x(\la) \abra{\tau}{\la}\xi^a\xi^b\xi^c\frac{\partial^3 \AAA(\la)}{\partial \chi^a \partial \chi^b\partial \chi^c}h_x(\la)\\ 
 &-3\int \DD\la\DD\la' h^{-1}_x(\la) \abra{\tau}{\la}\xi^a\xi^b\frac{\partial^2 \AAA(\la)}{\partial\chi^a\partial\chi^b}\frac{U_x(\la,\la')}{\abra{\la}{\la'}}\xi^c\frac{\partial \AAA(\la')}{\partial\chi'^{c}}h_x(\la')\\
 &-3\int \DD\la\DD\la' h^{-1}_x(\la) \xi^a\frac{\partial \AAA(\la)}{\partial\chi^a}\frac{U_x(\la,\la')}{\abra{\la}{\la'}}\abra{\tau}{\la'}\xi^b\xi^c\frac{\partial^2 \AAA(\la')}{\partial\chi'^{b}\partial\chi'^{c}}h_x(\la')\\
 &+6\int \DD\la\DD\la'\DD\la'' h^{-1}_x(\la) \xi^a\frac{\partial \AAA(\la)}{\partial\chi^{a}}\frac{U_x(\la,\la')}{\abra{\la}{\la'}}\abra{\tau}{\la'}\xi^b\frac{\partial \AAA(\la')}{\partial\chi'^{b}}\frac{U_x(\la',\la'')}{\abra{\la'}{\la''}}\xi^c\frac{\partial \AAA(\la'')}{\partial\chi''^{c}}h_x(\la'')
 \eqndot
\end{aligned}
\end{equation}
\allowdisplaybreaks
The vertex of the self-dual part of the field strength $F=\tau^\alpha\tau^\beta\xi^a\xi^b\xi^c\xi^dF_{\alpha\beta abcd}(x)$ equals
\begin{align}
\nonumber
 &\mathbf{W}_{F}= \int \DD\la h^{-1} \abra{\tau}{\la}^2\xi^a\xi^b\xi^c\xi^d\frac{\partial^4 \AAA(\la)}{\partial \chi^a \partial \chi^b\partial \chi^c \partial \chi^d}h\\ \nonumber
 &+6\int \DD\la\DD\la'h^{-1} \abra{\tau}{\la}\xi^a\xi^b\frac{\partial^2 \AAA(\la)}{\partial\chi^a\partial\chi^b}\frac{U_x(\la,\la')}{\abra{\la}{\la'}}\abra{\tau}{\la'}\xi^c\xi^d\frac{\partial^2 \AAA(\la')}{\partial\chi'^c\partial\chi'^{d}}h'\\ \nonumber
 &-4\int \DD\la\DD\la'h^{-1} \abra{\tau}{\la}^2\xi^a\xi^b\xi^c\frac{\partial^3 \AAA(\la)}{\partial\chi^a\partial\chi^b\partial\chi^c}\frac{U_x(\la,\la')}{\abra{\la}{\la'}}\xi^d\frac{\partial \AAA(\la')}{\partial\chi'^{d}}h'\\ \nonumber
 &-4\int \DD\la\DD\la'h^{-1} \xi^a \frac{\partial \AAA(\la)}{\partial\chi^a}\frac{U_x(\la,\la')}{\abra{\la}{\la'}}\abra{\tau}{\la'}^2\xi^b\xi^c\xi^d\frac{\partial^3 \AAA(\la')}{\partial\chi'^{b}\partial\chi'^{c}\partial\chi'^{d}}h'\\ \nonumber
 &-12\int \DD\la\DD\la'\DD\la''h^{-1} \abra{\tau}{\la}\xi^a\xi^b\la^{\alpha}\frac{\partial^2 \AAA(\la)}{\partial\chi^{a}\partial\chi^{b}}\frac{U_x(\la,\la')}{\abra{\la}{\la'}}\abra{\tau}{\la'}\xi^c\frac{\partial \AAA(\la')}{\partial\chi'^{c}}\frac{U_x(\la',\la'')}{\abra{\la'}{\la''}}\xi^d\frac{\partial \AAA(\la'')}{\partial\chi''^{d}}h''\\ \nonumber
 &-12\int \DD\la\DD\la'\DD\la''h^{-1} \xi^a\frac{\partial \AAA(\la)}{\partial\chi^{a}}\frac{U_x(\la,\la')}{\abra{\la}{\la'}}\abra{\tau}{\la'}^2\xi^b\xi^c\frac{\partial^2 \AAA(\la')}{\partial\chi'^{b}\partial\chi'^{c}}\frac{U_x(\la',\la'')}{\abra{\la'}{\la''}}\xi^d\frac{\partial \AAA(\la'')}{\partial\chi''^{d}}h''\\ \nonumber
 &-12\int \DD\la\DD\la'\DD\la''h^{-1} \xi^a\frac{\partial \AAA(\la)}{\partial\chi^{a}}\frac{U_x(\la,\la')}{\abra{\la}{\la'}}\abra{\tau}{\la'}\xi^b\frac{\partial \AAA(\la')}{\partial\chi'^{b}}\frac{U_x(\la',\la'')}{\abra{\la'}{\la''}}\abra{\tau}{\la''}\xi^c\xi^d\frac{\partial^2 \AAA(\la'')}{\partial\chi''^{c}\partial\chi''^{d}}h''\\ \nonumber
 &+24\int \DD\la\DD\la'\DD\la''\DD\la''' h^{-1} \xi^a\frac{\partial \AAA(\la)}{\partial\chi^{a}}\frac{U_x(\la,\la')}{\abra{\la}{\la'}}\abra{\tau}{\la'}\xi^b\frac{\partial \AAA(\la')}{\partial\chi'^{b}}\frac{U_x(\la',\la'')}{\abra{\la'}{\la''}}\abra{\tau}{\la''}\\&\qquad \qquad \times \xi^c\frac{\partial \AAA(\la'')}{\partial\chi''^{c}}\frac{U_x(\la'',\la''')}{\abra{\la''}{\la'''}}\xi^d\frac{\partial \AAA(\la''')}{\partial\chi'''^{d}}h'''
 \eqncom
 \label{eq: sdfieldstrength}
\end{align}
where we abbreviated $h\equiv h_x(\lambda)$, $h'\equiv h_x(\lambda')$ and so on.
For a scalar with covariant derivative $D\phi=-\tau^\alpha\bar{\tau}^{\dot\alpha}\xi^a\xi^bD_{\alpha\dot\alpha}\phi_{ab}$, we find
\begin{equation}
\label{eq:covariantderivativephi}
\begin{aligned}
 \mathbf{W}_{D\phi}= &-\int \DD\la h^{-1} \abra{\tau}{\la}\bar{\tau}^{\dot\alpha}\xi^a\xi^b\frac{\partial^3 \AAA(\la)}{\partial\mu^{\dot\alpha}\partial\chi^a\partial\chi^b}h\\
 &+2\int \DD\la\DD\la' h^{-1} \abra{\tau}{\la}\bar{\tau}^{\dot\alpha}\xi^a\frac{\partial^2 \AAA(\la)}{\partial\mu^{\dot\alpha}\partial\chi^a}\frac{U_x(\la,\la')}{\abra{\la}{\la'}}\xi^b\frac{\partial \AAA(\la')}{\partial\chi'^{b}}h'\\
 &+2\int \DD\la\DD\la' h^{-1} \xi^a\frac{\partial \AAA(\la)}{\partial\chi^a}\frac{U_x(\la,\la')}{\abra{\la}{\la'}}\abra{\tau}{\la'}\bar{\tau}^{\dot\alpha}\xi^b\frac{\partial^2 \AAA(\la')}{\partial\mu'^{\dot\alpha}\partial\chi'^{b}}h'\\
 &-\int \DD\la\DD\la' h^{-1} \abra{\tau}{\la}\xi^a\xi^b\frac{\partial^2 \AAA(\la)}{\partial\chi^a\partial\chi^{b}}\frac{U_x(\la,\la')}{\abra{\la}{\la'}}\bar{\tau}^{\dot\alpha}\frac{\partial \AAA(\la')}{\partial\mu'^{\dot\alpha}}h'\\
 &-\int \DD\la\DD\la' h^{-1} \bar{\tau}^{\dot\alpha}\frac{\partial \AAA(\la)}{\partial\mu^{\dot\alpha}}\frac{U_x(\la,\la')}{\abra{\la}{\la'}}\abra{\tau}{\la'}\xi^a\xi^b\frac{\partial^2 \AAA(\la')}{\partial\chi'^a\partial\chi'^{b}}h'\\
 &+2\int \DD\la\DD\la'\DD\la'' h^{-1} 
 \bar{\tau}^{\dot\alpha}\frac{\partial \AAA(\la)}{\partial\mu^{\dot\alpha}}\frac{U_x(\la,\la')}{\abra{\la}{\la'}}\abra{\tau}{\la'}\xi^a\frac{\partial \AAA(\la')}{\partial\chi'^a}\frac{U_x(\la',\la'')}{\abra{\la'}{\la''}}\xi^b\frac{\partial \AAA(\la'')}{\partial\chi''^b}h''\\
 &+2\int \DD\la\DD\la'\DD\la'' h^{-1} \xi^a\frac{\partial \AAA(\la)}{\partial\chi^a}\frac{U_x(\la,\la')}{\abra{\la}{\la'}}\abra{\tau}{\la'}\bar{\tau}^{\dot\alpha}\frac{\partial \AAA(\la')}{\partial\mu'^{\dot\alpha}}\frac{U_x(\la',\la'')}{\abra{\la'}{\la''}}\xi^b\frac{\partial \AAA(\la'')}{\partial\chi''^b}h''\\
 &+2\int \DD\la\DD\la'\DD\la'' h^{-1} \xi^a\frac{\partial \AAA(\la)}{\partial\chi^{a}}\frac{U_x(\la,\la')}{\abra{\la}{\la'}}\abra{\tau}{\la'}\xi^b\frac{\partial \AAA(\la')}{\partial\chi'^b}\frac{U_x(\la',\la'')}{\abra{\la'}{\la''}}\bar{\tau}^{\dot\alpha}\frac{\partial \AAA(\la'')}{\partial\mu''^{\dot\alpha}}h''
 \eqndot
\end{aligned}
\end{equation}
In all the above expressions, it is understood that $\calA(\la)\equiv\calA(\calZ_x(\la))$ and so on.

\section{Derivation of the MHV form factors}
\label{app: derivation}

In this appendix, we derive our result \eqref{eq: general MHV form factor} for the tree-level $n$-point MHV form factors of all composite operators using the second strategy sketched in section \ref{subsec: all MHV form factors} of the main text.  
To this end,  we take our cogwheel Wilson loop \eqref{eq:finaldefinitionWilsonloop} with $3L$ edges. 
For each term in \eqref{eq:finaldefinitionWilsonloop}, we use the expression \eqref{eq:frameUdefinitionpart1} for the parallel propagators $U$.
Let $\calO=\Tr(D^{k_1}A_1\cdots D^{k_L}A_L)$ be our local operator with fields defined in \eqref{eq: alphabet of fields with polarizations} and let $n$ be the total number of external on-shell fields.

\begin{figure}[htbp]
 \centering
  \includegraphics[height=4.5cm]{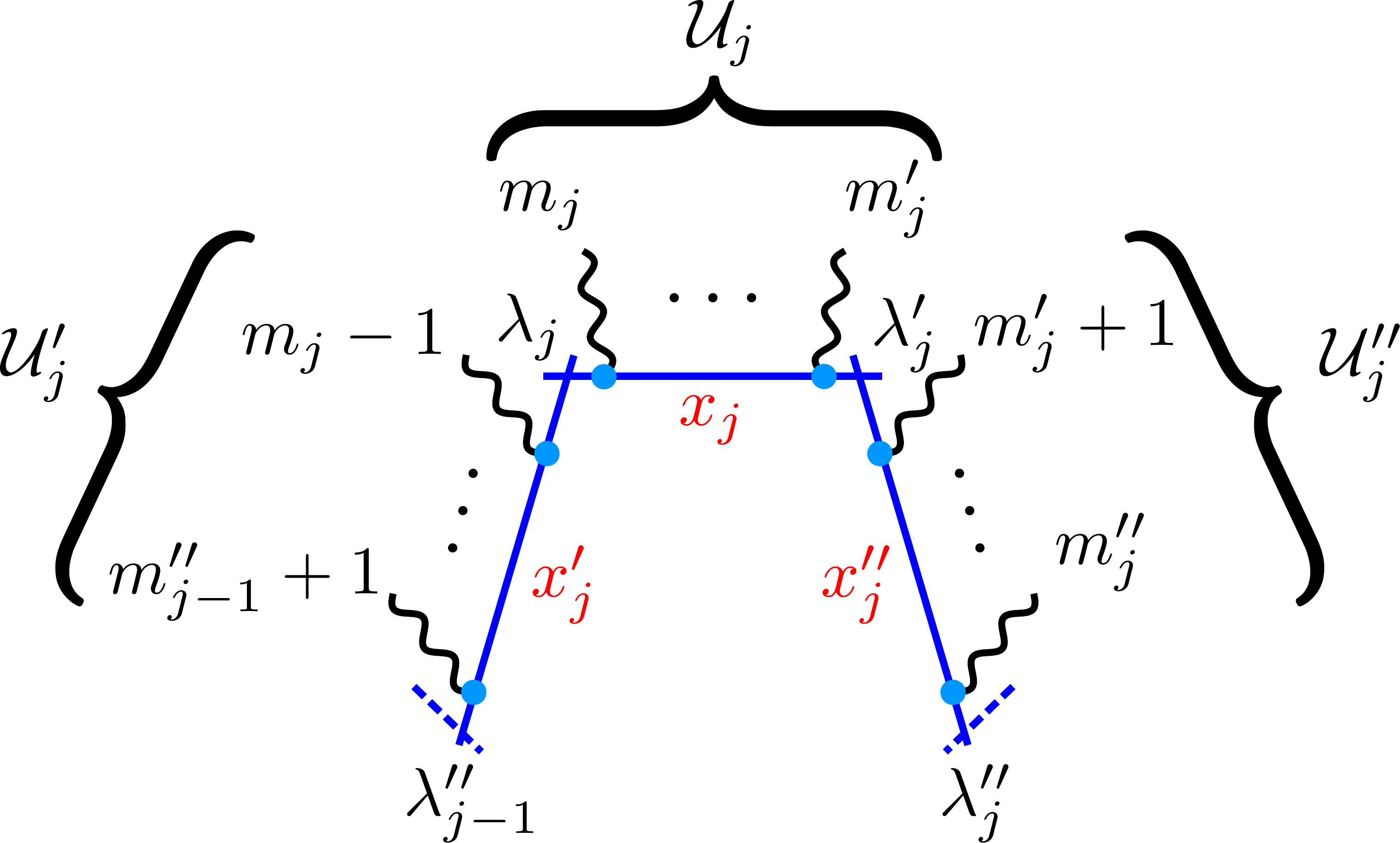}
  \caption{\it This figure shows the labels of the on-shell fields attached to each tooth of the cogwheel Wilson loop. Since the operator bearing edge $x_j$ must have at least one external field attached, we have the cyclic constraints $\cdots \leq m_{j-1}''< m_j\leq m_j'\leq m_j''< \cdots $. We denote by $\mathcal{U}_j,\cdots$ the contributions to the form factors from the different sides of the cog.
  }
  \label{fig:CogwheelPiece}
\end{figure}

We consider each cog (or tooth) of the cogwheel Wilson loop separately, see figure~\ref{fig:CogwheelPiece}.  We first look at the edges of the loop that carry the irreducible fields $D^{k_j}A_j$ of $\calO$, i.e.\ at the parallel propagators $U_{x_j}(\la_j,\la_j')$.  
We take the term in \eqref{eq:frameUdefinitionpart1} with $(m_j'-m_j+1)$ $\calA$'s and insert external on-shell fields labeled $m_j,\dots,m_j'$ into it. We must have $m_j\leq m_j'$, since we want to emit at least one on-shell particle from the edges that carry the irreducible fields; otherwise, the derivatives yield zero. 

The calculation is similar to those previously employed in \eqref{eq:MHV amplitude} and \eqref{eq:deltaintegration}, and it yields
\begin{align}
 \mathcal{U}_j(m_j,m_j')&=\int_{}\frac{\abra{\lambda_j}{\lambda'_{j}}\DD\tilde{\la}_1 \cdots  \DD\tilde{\la}_{m_j'-m_j+1}}{\abra{\lambda_j}{\tilde{\la}_1}\abra{\tilde{\la}_1}{\tilde{\la}_2}\cdots \abra{\tilde{\la}_{m_j'-m_j+1}}{\lambda_{j}'}}\AAA_{\fP_{m_j}}(\calZ_{x_j}(\tilde{\la}_1)) \cdots \AAA_{\fP_{m_j'}}(\calZ_{x_j}(\tilde{\la}_{m_j'-m_j+1}))\notag\\
&= \frac{\abra{ \la_j}{\la_{j}'}}{\abra{\lambda_j}{p_{m_j}}\prod_{k=m_j}^{m_j'-1}\abra{p_k}{p_{k+1}}\abra{p_{m_j'}}{\lambda_j'}}\e^{ i\sum_{k=m_j}^{m_j'}(x_j\fp_k+\theta_j p_k\eta_k)}\,.
\end{align}
We then act on it with the forming operator \eqref{eq:definitionformingfactoronshellstates} for the irreducible field $D^{k_j}A_j$, take the operator limit \eqref{eq:oplimit1}, \eqref{eq:oplimit2} and set $\theta=0$:
\begin{equation}
 \label{eq: formingfactoronshellstates on wilson line}
\begin{multlined}
\mathcal{I}_j(m_j,m_j')=\lim_{\hexagon\rightarrow \xdot}\PPP_{D^{k_j}A_j} \mathcal{U}_j(m_j,m_j')_{\big{|}\theta=0}\,,
\end{multlined}
\end{equation}
which we write explicitly as
\begin{equation}
 \label{eq: formingfactoronshellstates on wilson line operator limit}
\begin{multlined}
\mathcal{I}_j(m_j,m_j')=\frac{-1}{\abra{\tau_j}{p_{m_j}}\prod_{k=m_j}^{m_j'-1}\abra{p_k}{p_{k+1}}\abra{p_{m_j'}}{\tau_j}}
\Bigg(\sum_{k=m_j}^{m_j'}\abra{\tau_j}{p_k}\sbra{\bar{p}_k}{\bar{\tau}_j}\Bigg)^{k_i}\\
\left\{\begin{array}{ll}
\big(\sum_{k=m_j}^{m_j'}\abra{\tau_j}{p_k}\sbra{\bar{p}_k}{\bar{\tau}_j}\big)^2 & \text{ for } A_j=\bar F\\
\big(\sum_{k=m_j}^{m_j'}\abra{\tau_j}{p_k}\sbra{\bar{p}_k}{\bar{\tau}_j}\big)\big(\sum_{k=m_j}^{m_j'}\abra{\tau_j}{p_k}\{\xi_j\eta_k\}\big)& \text{ for }A_j=\bar{\psi}\\
\big(\sum_{k=m_j}^{m_j'}\abra{\tau_j}{p_k}\{\xi_j\eta_k\}\big)^2& \text{ for } A_j=\phi\\
\big(\sum_{k=m_j}^{m_j'}\abra{\tau_j}{p_k}\{\xi_j\eta_k\}\big)^3&\text{ for }A_j=\psi\\
\big(\sum_{k=m_j}^{m_j'}\abra{\tau_j}{p_k}\{\xi_j\eta_k\}\big)^4& \text{ for } A_j=F
\end{array}\right\} \e^{ i\sum_{k=m_j}^{m_j'}x\fp_k} \eqndot
\end{multlined}
\end{equation}
Recalling the notation introduced in subsection \ref{subsec: all MHV form factors}, we can write \eqref{eq: formingfactoronshellstates on wilson line operator limit} as
\beq
\label{eq: formingfactoronshellstates on wilson line operator limit 2}
\mathcal{I}_j(m_j,m_j')=-\frac{\left(\sum_{k=m_j}^{m_j'}\abra{\tau_j}{p_k}\sbra{\bar{p}_k}{\bar{\tau}_j}\right)^{N_j-n_{\theta_j}}\left(\sum_{k=m_j}^{m_j'}\abra{\tau_j}{p_k}\{\xi_j\eta_k\}\right)^{n_{\theta_j}}}{\abra{\tau_j}{p_{m_j}}\prod_{k=m_j}^{m_j'-1}\abra{p_k}{p_{k+1}}\abra{p_{m_j'}}{\tau_j}}\e^{ i\sum_{k=m_j}^{m_j'}x\fp_k}\,.
\eeq
Using a slight generalization of the identity
\beq
\left(\sum_{k=1}^m\mathcal{C}_k\right)^N=\sum_{1\leq k_1\leq k_2\leq \cdots \leq k_N\leq m}\frac{N!}{M(\{k_1,\ldots, k_N\})!}\mathcal{C}_{k_1}\cdots \mathcal{C}_{k_N}\,,
\eeq
we can rewrite \eqref{eq: formingfactoronshellstates on wilson line operator limit 2} as 
\beq
\begin{aligned}
\label{eq: formingfactoronshellstates on wilson line operator limit 3}
\mathcal{I}_j(m_j,m_j')&=-\frac{\e^{ i\sum_{k=m_j}^{m_j'}x\fp_k}}{\abra{\tau_j}{p_{m_j}}\prod_{k=m_j}^{m_j'-1}\abra{p_k}{p_{k+1}}\abra{p_{m_j'}}{\tau_j}}\sum_{m_j\leq B_{j,1}\leq \cdots \leq B_{j,N_j}\leq m_j'}
\left(\prod_{k=1}^{N_j}\abra{\tau_j}{p_{\notA_{j,k}}}\right)\\
  &\phaneq\phantom{-}\times\sum_{\sigma\in S_{N_j}}
  \frac{\{\xi_j \eta_{\sigma(\notA_{j,1})}\}\cdots\{\xi_j \eta_{\sigma(\notA_{j,n_{\theta_j}})}\}\sbra{\bar{p}_{\sigma(\notA_{j,n_{\theta_j}+1})}}{\bar\tau_j}\cdots\sbra{\bar{p}_{\sigma(\notA_{j,N_j})}}{\bar\tau_j}}{M(\{\notA_{j,1},\dots,\notA_{j,N_j}\})!}
\,.
\end{aligned}
\eeq

In addition to the factor \eqref{eq: formingfactoronshellstates on wilson line operator limit 3}, a contribution from the two edges of the Wilson loop on the left and right of $x_j$ occurs, which are not acted on by derivative operators. 
Similarly to $\mathcal{U}_j$, we can compute the contributions from the two other sides of the cogwheel tooth, see figure~\ref{fig:CogwheelPiece}. Specifically, we find
\beq
\mathcal{U}_j'(m_{j-1}'',m_j)= \frac{\abra{ \la_{j-1}''}{\la_{j}}}{\abra{\lambda_{j-1}''}{p_{m_{j-1}''+1}}\prod_{k=m_{j-1}''+1}^{m_j-2}\abra{p_k}{p_{k+1}}\abra{p_{m_j-1}}{\lambda_j}}\e^{ i\sum_{k=m_{j-1}''+1}^{m_j-1}x_j'\fp_k}
\eeq
if $m_{j-1}''<m_j-1$ and $\mathcal{U}_j'=1$ if $m_{j-1}''=m_j-1$. Finally, we get
\beq
\mathcal{U}_j''(m_j',m_j'')= \frac{\abra{ \la_j'}{\la_{j}''}}{\abra{\lambda_j'}{p_{m_j'+1}}\prod_{k=m_j'+1}^{m_j''-1}\abra{p_k}{p_{k+1}}\abra{p_{m_j''}}{\lambda_{j}''}}\e^{ i\sum_{k=m_j'+1}^{m_j''}x_j''\fp_k}
\eeq
if $m_j'<m_j''$ and $\mathcal{U}_j''=1$ if $m_j'=m_j''$.
After taking the operator limit \eqref{eq:oplimit1} and \eqref{eq:oplimit2}, these yield
\beq
\label{eq: empty wilson line operator limit 1}
\mathcal{I}_j'(m_{j-1}'',m_j) =\frac{\abra{ \n}{\tau_{j}}}{\abra{\n}{p_{m_{j-1}''+1}}\prod_{k=m_{j-1}''+1}^{m_j-2}\abra{p_k}{p_{k+1}}\abra{p_{m_j-1}}{\tau_j}}\e^{ i\sum_{k=m_{j-1}''+1}^{m_j-1}x\fp_k}
\eeq
as well as 
\beq
\label{eq: empty wilson line operator limit 2}
\mathcal{I}_j''(m_j',m_j'')= \frac{\abra{ \tau_j}{\n}}{\abra{\tau_j}{p_{m_j'+1}}\prod_{k=m_j'+1}^{m_j''-1}\abra{p_k}{p_{k+1}}\abra{p_{m_j''}}{\n}}\e^{ i\sum_{k=m_j'+1}^{m_j''}x\fp_k}
\eqndot
\eeq
As for $\mathcal{U}_j'$ and $\mathcal{U}_j''$, we have by definition that $\mathcal{I}_j'(m_{j-1}'',m_j)=1$ if $m_{j-1}''=m_j-1$ and that $\mathcal{I}_j''(m_j',m_j'')=1$ if $m_j'=m_j''$. 
The position-space form factor of the operator $\calO$ is now obtained by taking the product of all \eqref{eq: formingfactoronshellstates on wilson line operator limit 2}, \eqref{eq: empty wilson line operator limit 1} and \eqref{eq: empty wilson line operator limit 2} with $1\leq j\leq L$ and then summing over all possible indices $m_j$, $m_j'$ and $m_j''$. Since the irreducible fields are placed on the edges $x_j$, we see from figure~\ref{fig:CogwheelPiece} that we can have $\cdots \leq m_{j-1}''< m_j\leq m_j'\leq m_j''< \cdots $. Hence, the sum is cyclic over 
$1\leq m_1\leq m_1'\leq m_1''< m_2\leq m_2'\leq m_2''<m_3\leq \cdots <m_L\leq m_L'\leq m_L''=m_1+n-1$
 and we write for the form factor:
\begin{equation}
\label{eq: general MHV form factor in position space}
\calF_{\calO}(1,\dots,n;x)=\sum_{\{m_j,m_j',m_j''\}} \prod_{j=1}^L \mathcal{I}_j'(m_{j-1}'',m_j) \mathcal{I}_j(m_j,m_j')\mathcal{I}_j''(m_j',m_j'')\,.
\end{equation}
Let us now for simplicity denote by $\tilde{\mathcal{I}}_j$, $\tilde{\mathcal{I}}_j'$ and $\tilde{\mathcal{I}}_j''$ the contributions \eqref{eq: formingfactoronshellstates on wilson line operator limit 2}, \eqref{eq: empty wilson line operator limit 1} and \eqref{eq: empty wilson line operator limit 2} stripped off the exponential factors. We can almost immediately perform the sum over the $m_j$, $m_j'$ and $m_j''$, leaving in \eqref{eq: general MHV form factor in position space} only the sums over the $B_{i,j}$ that are contained implicitly in the $\mathcal{I}_j$. In order to do that, it turns out to be useful to rewrite the MHV prefactor of \eqref{eq: formingfactoronshellstates on wilson line operator limit 3} as
\begin{align}
\frac{1}{\abra{\tau_j}{p_{m_j}}\prod_{k=m_j}^{m_j'-1}\abra{p_k}{p_{k+1}}\abra{p_{m_j'}}{\tau_j}}=&\frac{\abra{\tau_j}{p_{B_{j,1}}}}{\abra{\tau_j}{p_{m_j}}\prod_{k=m_j}^{B_{j,1}-1}\abra{p_k}{p_{k+1}}}\frac{1}{\abra{\tau_j}{p_{B_{j,1}}}\prod_{k=B_{j,1}}^{B_{j,N_j}-1}\abra{p_k}{p_{k+1}}\abra{p_{B_{j,N_j}}}{\tau_j}}\notag\\&\times \frac{\abra{p_{B_{j,N_j}}}{\tau_j}}{\prod_{k=B_{j,N_j}}\abra{p_k}{p_{k+1}}\abra{p_{m_j'}}{\tau_j}}\,.
\label{eq: denominator decomposition}
\end{align}
\begin{figure}[tbp]
 \centering
  \includegraphics[height=3.5cm]{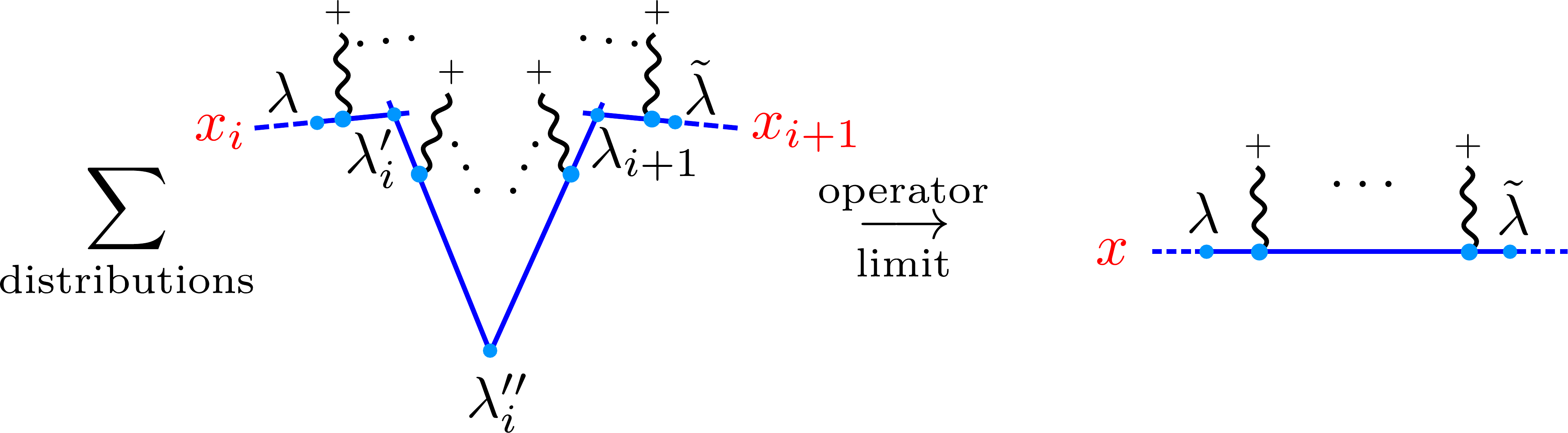}
  \caption{\it The contributions to the emissions of positive helicity gluons from the different edges combine in the operator limit.
  }
  \label{fig:CogwheelGaugeInvariance}
\end{figure}
Using multiple times the following telescopic Schouten identity
\beq
\label{eq:telescopedSchouten}
\frac{\abra{a}{b}}{\abra{a}{1}\cdots \abra{m}{b}}+\sum_{k=1}^{m-1}\frac{\abra{a}{b}\abra{b}{c}}{\abra{a}{1}\cdots \abra{k}{b}\abra{b}{(k+1)}\cdots \abra{m}{c}}+\frac{\abra{b}{c}}{\abra{b}{1}\cdots \abra{m}{c}}=\frac{\abra{a}{c}}{\abra{a}{1}\cdots \abra{m}{c}}\,,
\eeq 
we can then show the identity
\begin{multline}
\label{eq:sumoversidesofcogs}
\sum_{B_{j,N_j}\leq m_j'\leq m_j''<m_{j+1}\leq B_{j+1,1}}\frac{\abra{p_{B_{j,N_j}}}{\tau_j}\tilde{\mathcal{I}}_j''(m_j',m_j'')\tilde{\mathcal{I}}_{j+1}'(m_{j}'',m_{j+1})\abra{\tau_{j+1}}{p_{B_{j+1,1}}}}{\prod_{k=B_{j,N_j}}\abra{p_k}{p_{k+1}}\abra{p_{m_j'}}{\tau_j}\abra{\tau_{j+1}}{p_{m_{j+1}}}\prod_{k=m_{j+1}}^{B_{j+1,1}-1}\abra{p_k}{p_{k+1}}}
\\=\frac{\abra{p_{B_{j,N_j}}}{p_{B_{j+1,1}}}}{\prod_{k=B_{j,N_j}}^{B_{j+1,1}-1}\abra{p_k}{p_{k+1}}}\,.
\end{multline}
Thus, the auxiliary spinor $\n$ in $\tilde{\mathcal{I}}_j'$ and $\tilde{\mathcal{I}}_{j}''$, which could not be part of our final result, drops out.
 In fact, the identity \eqref{eq:sumoversidesofcogs} is the direct consequence of the following identity for the parallel propagators $U_{x}(\lambda,\lambda_j')U_{x}(\lambda_j',\lambda_j'')U_{x}(\lambda_j'',\lambda_{j+1})U_{x}(\lambda_{j+1},\tilde\lambda)=U_{x}(\lambda,\tilde\lambda)$, see figure~\ref{fig:CogwheelGaugeInvariance}, which holds after taking the operator limit.

Hence, using \eqref{eq: general MHV form factor in position space}, the expression \eqref{eq: formingfactoronshellstates on wilson line operator limit 3} and the identity \eqref{eq:sumoversidesofcogs}, we obtain after Fourier transforming the claimed result \eqref{eq: general MHV form factor}.
In particular, the first and the last term in the product at the end of the first line of \eqref{eq: formingfactoronshellstates on wilson line operator limit 3} always cancel with corresponding terms in the denominator on the right hand side of \eqref{eq: denominator decomposition} and the global sign in \eqref{eq: formingfactoronshellstates on wilson line operator limit 3}.


\bibliographystyle{../JHEP}
\bibliography{../biblio}

\providecommand{\href}[2]{#2}\begingroup\raggedright\begin{thebibliography}{10}

\bibitem{Koster:2016ebi}
L.~Koster, V.~Mitev, M.~Staudacher, and M.~Wilhelm, {\it {Composite Operators
  in the Twistor Formulation of $\mathcal{N}=4$ SYM Theory}},
  \href{http://arxiv.org/abs/1603.04471}{{\tt arXiv:1603.04471}}.

\bibitem{Beisert:2010jr}
N.~Beisert {\em et~al.}, {\it {Review of AdS/CFT Integrability: An Overview}},
  {\em Lett. Math. Phys.} {\bf 99} (2012) 3--32,
  [\href{http://arxiv.org/abs/1012.3982}{{\tt arXiv:1012.3982}}].

\bibitem{Elvang:2013cua}
H.~Elvang and Y.-t. Huang, {\it {Scattering Amplitudes}},
  \href{http://arxiv.org/abs/1308.1697}{{\tt arXiv:1308.1697}}.

\bibitem{Henn:2014yza}
J.~M. Henn and J.~C. Plefka, {\it {Scattering Amplitudes in Gauge Theories}},
  {\em Lect. Notes Phys.} {\bf 883} (2014) 1--195.

\bibitem{Boels:2006ir}
R.~Boels, L.~Mason, and D.~Skinner, {\it {Supersymmetric Gauge Theories in
  Twistor Space}},  {\em JHEP} {\bf 0702} (2007) 014,
  [\href{http://arxiv.org/abs/hep-th/0604040}{{\tt hep-th/0604040}}].

\bibitem{Adamo:2011pv}
T.~Adamo, M.~Bullimore, L.~Mason, and D.~Skinner, {\it {Scattering Amplitudes
  and Wilson Loops in Twistor Space}},  {\em J.Phys.} {\bf A44} (2011) 454008,
  [\href{http://arxiv.org/abs/1104.2890}{{\tt arXiv:1104.2890}}].

\bibitem{Boels:2007qn}
R.~Boels, L.~J. Mason, and D.~Skinner, {\it {From twistor actions to MHV
  diagrams}},  {\em Phys. Lett.} {\bf B648} (2007) 90--96,
  [\href{http://arxiv.org/abs/hep-th/0702035}{{\tt hep-th/0702035}}].

\bibitem{Adamo:2011cb}
T.~Adamo and L.~Mason, {\it {MHV diagrams in twistor space and the twistor
  action}},  {\em Phys.Rev.} {\bf D86} (2012) 065019,
  [\href{http://arxiv.org/abs/1103.1352}{{\tt arXiv:1103.1352}}].

\bibitem{Mason:2010yk}
L.~J. Mason and D.~Skinner, {\it {The Complete Planar S-matrix of
  $\mathcal{N}=4$ SYM as a Wilson Loop in Twistor Space}},  {\em JHEP} {\bf 12}
  (2010) 018, [\href{http://arxiv.org/abs/1009.2225}{{\tt arXiv:1009.2225}}].

\bibitem{Bullimore:2011ni}
M.~Bullimore and D.~Skinner, {\it {Holomorphic Linking, Loop Equations and
  Scattering Amplitudes in Twistor Space}},
  \href{http://arxiv.org/abs/1101.1329}{{\tt arXiv:1101.1329}}.

\bibitem{ArkaniHamed:2009dn}
N.~Arkani-Hamed, F.~Cachazo, C.~Cheung, and J.~Kaplan, {\it {A Duality For The
  S Matrix}},  {\em JHEP} {\bf 03} (2010) 020,
  [\href{http://arxiv.org/abs/0907.5418}{{\tt arXiv:0907.5418}}].

\bibitem{Mason:2009qx}
L.~J. Mason and D.~Skinner, {\it {Dual Superconformal Invariance, Momentum
  Twistors and Grassmannians}},  {\em JHEP} {\bf 11} (2009) 045,
  [\href{http://arxiv.org/abs/0909.0250}{{\tt arXiv:0909.0250}}].

\bibitem{Bullimore:2010pj}
M.~Bullimore, L.~J. Mason, and D.~Skinner, {\it {MHV Diagrams in Momentum
  Twistor Space}},  {\em JHEP} {\bf 12} (2010) 032,
  [\href{http://arxiv.org/abs/1009.1854}{{\tt arXiv:1009.1854}}].

\bibitem{CaronHuot:2010ek}
S.~Caron-Huot, {\it {Notes on the scattering amplitude / Wilson loop duality}},
   {\em JHEP} {\bf 07} (2011) 058, [\href{http://arxiv.org/abs/1010.1167}{{\tt
  arXiv:1010.1167}}].

\bibitem{Belitsky:2011zm}
A.~V. Belitsky, G.~P. Korchemsky, and E.~Sokatchev, {\it {Are scattering
  amplitudes dual to super Wilson loops?}},  {\em Nucl. Phys.} {\bf B855}
  (2012) 333--360, [\href{http://arxiv.org/abs/1103.3008}{{\tt
  arXiv:1103.3008}}].

\bibitem{Koster:2014fva}
L.~Koster, V.~Mitev, and M.~Staudacher, {\it {A Twistorial Approach to
  Integrability in $\mathcal N=$ 4 SYM}},  {\em Fortsch. Phys.} {\bf 63}
  (2015), no.~2 142--147, [\href{http://arxiv.org/abs/1410.6310}{{\tt
  arXiv:1410.6310}}].

\bibitem{Chicherin:2014uca}
D.~Chicherin, R.~Doobary, B.~Eden, P.~Heslop, G.~P. Korchemsky, L.~Mason, and
  E.~Sokatchev, {\it {Correlation functions of the chiral stress-tensor
  multiplet in $ \mathcal{N}=4 $ SYM}},  {\em JHEP} {\bf 06} (2015) 198,
  [\href{http://arxiv.org/abs/1412.8718}{{\tt arXiv:1412.8718}}].

\bibitem{vanNeerven:1985ja}
W.~L. van Neerven, {\it {Infrared Behavior of On-shell Form-factors in a
  $\mathcal{N}=4$ Supersymmetric {Yang-Mills} Field Theory}},  {\em Z. Phys.}
  {\bf C30} (1986) 595.

\bibitem{Brandhuber:2010ad}
A.~Brandhuber, B.~Spence, G.~Travaglini, and G.~Yang, {\it {Form Factors in
  $\mathcal{N}=4$ Super Yang-Mills and Periodic Wilson Loops}},  {\em JHEP}
  {\bf 1101} (2011) 134, [\href{http://arxiv.org/abs/1011.1899}{{\tt
  arXiv:1011.1899}}].

\bibitem{Bork:2010wf}
L.~V. Bork, D.~I. Kazakov, and G.~S. Vartanov, {\it {On form factors in
  $\mathcal{N}=4$ sym}},  {\em JHEP} {\bf 02} (2011) 063,
  [\href{http://arxiv.org/abs/1011.2440}{{\tt arXiv:1011.2440}}].

\bibitem{Brandhuber:2011tv}
A.~Brandhuber, O.~Gurdogan, R.~Mooney, G.~Travaglini, and G.~Yang, {\it
  {Harmony of Super Form Factors}},  {\em JHEP} {\bf 10} (2011) 046,
  [\href{http://arxiv.org/abs/1107.5067}{{\tt arXiv:1107.5067}}].

\bibitem{Bork:2011cj}
L.~V. Bork, D.~I. Kazakov, and G.~S. Vartanov, {\it {On MHV Form Factors in
  Superspace for $\mathcal{N}=4$ SYM Theory}},  {\em JHEP} {\bf 10} (2011) 133,
  [\href{http://arxiv.org/abs/1107.5551}{{\tt arXiv:1107.5551}}].

\bibitem{Henn:2011by}
J.~M. Henn, S.~Moch, and S.~G. Naculich, {\it {Form factors and scattering
  amplitudes in $\mathcal{N}=4$ SYM in dimensional and massive
  regularizations}},  {\em JHEP} {\bf 12} (2011) 024,
  [\href{http://arxiv.org/abs/1109.5057}{{\tt arXiv:1109.5057}}].

\bibitem{Gehrmann:2011xn}
T.~Gehrmann, J.~M. Henn, and T.~Huber, {\it {The three-loop form factor in
  $\mathcal{N}=4$ super Yang-Mills}},  {\em JHEP} {\bf 03} (2012) 101,
  [\href{http://arxiv.org/abs/1112.4524}{{\tt arXiv:1112.4524}}].

\bibitem{Brandhuber:2012vm}
A.~Brandhuber, G.~Travaglini, and G.~Yang, {\it {Analytic two-loop form factors
  in $\mathcal{N}=4$ SYM}},  {\em JHEP} {\bf 05} (2012) 082,
  [\href{http://arxiv.org/abs/1201.4170}{{\tt arXiv:1201.4170}}].

\bibitem{Bork:2012tt}
L.~V. Bork, {\it {On NMHV form factors in $\mathcal{N}=4$ SYM theory from
  generalized unitarity}},  {\em JHEP} {\bf 01} (2013) 049,
  [\href{http://arxiv.org/abs/1203.2596}{{\tt arXiv:1203.2596}}].

\bibitem{Engelund:2012re}
O.~T. Engelund and R.~Roiban, {\it {Correlation functions of local composite
  operators from generalized unitarity}},  {\em JHEP} {\bf 1303} (2013) 172,
  [\href{http://arxiv.org/abs/1209.0227}{{\tt arXiv:1209.0227}}].

\bibitem{Johansson:2012zv}
H.~Johansson, D.~A. Kosower, and K.~J. Larsen, {\it {Two-Loop Maximal Unitarity
  with External Masses}},  {\em Phys. Rev.} {\bf D87} (2013), no.~2 025030,
  [\href{http://arxiv.org/abs/1208.1754}{{\tt arXiv:1208.1754}}].

\bibitem{Boels:2012ew}
R.~H. Boels, B.~A. Kniehl, O.~V. Tarasov, and G.~Yang, {\it {Color-kinematic
  Duality for Form Factors}},  {\em JHEP} {\bf 02} (2013) 063,
  [\href{http://arxiv.org/abs/1211.7028}{{\tt arXiv:1211.7028}}].

\bibitem{Penante:2014sza}
B.~Penante, B.~Spence, G.~Travaglini, and C.~Wen, {\it {On super form factors
  of half-BPS operators in $\mathcal{N}=4$ super Yang-Mills}},  {\em JHEP} {\bf
  04} (2014) 083, [\href{http://arxiv.org/abs/1402.1300}{{\tt
  arXiv:1402.1300}}].

\bibitem{Brandhuber:2014ica}
A.~Brandhuber, B.~Penante, G.~Travaglini, and C.~Wen, {\it {The last of the
  simple remainders}},  {\em JHEP} {\bf 08} (2014) 100,
  [\href{http://arxiv.org/abs/1406.1443}{{\tt arXiv:1406.1443}}].

\bibitem{Bork:2014eqa}
L.~V. Bork, {\it {On form factors in $ \mathcal{N}=4 $ SYM theory and
  polytopes}},  {\em JHEP} {\bf 12} (2014) 111,
  [\href{http://arxiv.org/abs/1407.5568}{{\tt arXiv:1407.5568}}].

\bibitem{Wilhelm:2014qua}
M.~Wilhelm, {\it {Amplitudes, Form Factors and the Dilatation Operator in
  $\mathcal{N}=4$ SYM Theory}},  {\em JHEP} {\bf 02} (2015) 149,
  [\href{http://arxiv.org/abs/1410.6309}{{\tt arXiv:1410.6309}}].

\bibitem{Nandan:2014oga}
D.~Nandan, C.~Sieg, M.~Wilhelm, and G.~Yang, {\it {Cutting through form factors
  and cross sections of non-protected operators in $ \mathcal{N}=4 $ SYM}},
  {\em JHEP} {\bf 06} (2015) 156, [\href{http://arxiv.org/abs/1410.8485}{{\tt
  arXiv:1410.8485}}].

\bibitem{Loebbert:2015ova}
F.~Loebbert, D.~Nandan, C.~Sieg, M.~Wilhelm, and G.~Yang, {\it {On-Shell
  Methods for the Two-Loop Dilatation Operator and Finite Remainders}},  {\em
  JHEP} {\bf 10} (2015) 012, [\href{http://arxiv.org/abs/1504.06323}{{\tt
  arXiv:1504.06323}}].

\bibitem{Bork:2015fla}
L.~V. Bork and A.~I. Onishchenko, {\it {On soft theorems and form factors in $
  \mathcal{N}=4 $ SYM theory}},  {\em JHEP} {\bf 12} (2015) 030,
  [\href{http://arxiv.org/abs/1506.07551}{{\tt arXiv:1506.07551}}].

\bibitem{Frassek:2015rka}
R.~Frassek, D.~Meidinger, D.~Nandan, and M.~Wilhelm, {\it {On-shell diagrams,
  Gra{\ss}mannians and integrability for form factors}},  {\em JHEP} {\bf 01}
  (2016) 182, [\href{http://arxiv.org/abs/1506.08192}{{\tt arXiv:1506.08192}}].

\bibitem{Boels:2015yna}
R.~Boels, B.~A. Kniehl, and G.~Yang, {\it {Master integrals for the four-loop
  Sudakov form factor}},  {\em Nucl. Phys.} {\bf B902} (2016) 387--414,
  [\href{http://arxiv.org/abs/1508.03717}{{\tt arXiv:1508.03717}}].

\bibitem{Huang:2016bmv}
R.~Huang, Q.~Jin, and B.~Feng, {\it {Form Factor and Boundary Contribution of
  Amplitude}},  \href{http://arxiv.org/abs/1601.06612}{{\tt arXiv:1601.06612}}.

\bibitem{Alday:2007he}
L.~F. Alday and J.~Maldacena, {\it {Comments on gluon scattering amplitudes via
  AdS/CFT}},  {\em JHEP} {\bf 11} (2007) 068,
  [\href{http://arxiv.org/abs/0710.1060}{{\tt arXiv:0710.1060}}].

\bibitem{Maldacena:2010kp}
J.~Maldacena and A.~Zhiboedov, {\it {Form factors at strong coupling via a
  Y-system}},  {\em JHEP} {\bf 11} (2010) 104,
  [\href{http://arxiv.org/abs/1009.1139}{{\tt arXiv:1009.1139}}].

\bibitem{Gao:2013dza}
Z.~Gao and G.~Yang, {\it {Y-system for form factors at strong coupling in
  $AdS_5$ and with multi-operator insertions in $AdS_3$}},  {\em JHEP} {\bf 06}
  (2013) 105, [\href{http://arxiv.org/abs/1303.2668}{{\tt arXiv:1303.2668}}].

\bibitem{Wilhelm:2016izi}
M.~Wilhelm, {\em {Form factors and the dilatation operator in $\mathcal{N}=4$
  super Yang-Mills theory and its deformations}}.
\newblock PhD thesis, 2016.
\newblock \href{http://arxiv.org/abs/1603.01145}{{\tt arXiv:1603.01145}}.

\bibitem{Drummond:2008cr}
J.~M. Drummond and J.~M. Henn, {\it {All tree-level amplitudes in
  $\mathcal{N}=4$ SYM}},  {\em JHEP} {\bf 04} (2009) 018,
  [\href{http://arxiv.org/abs/0808.2475}{{\tt arXiv:0808.2475}}].

\bibitem{ArkaniHamed:2010kv}
N.~Arkani-Hamed, J.~L. Bourjaily, F.~Cachazo, S.~Caron-Huot, and J.~Trnka, {\it
  {The All-Loop Integrand For Scattering Amplitudes in Planar $\mathcal{N}=4$
  SYM}},  {\em JHEP} {\bf 01} (2011) 041,
  [\href{http://arxiv.org/abs/1008.2958}{{\tt arXiv:1008.2958}}].

\bibitem{Witten:2003nn}
E.~Witten, {\it {Perturbative gauge theory as a string theory in twistor
  space}},  {\em Commun.Math.Phys.} {\bf 252} (2004) 189--258,
  [\href{http://arxiv.org/abs/hep-th/0312171}{{\tt hep-th/0312171}}].

\bibitem{Cachazo:2004kj}
F.~Cachazo, P.~Svrcek, and E.~Witten, {\it {MHV vertices and tree amplitudes in
  gauge theory}},  {\em JHEP} {\bf 0409} (2004) 006,
  [\href{http://arxiv.org/abs/hep-th/0403047}{{\tt hep-th/0403047}}].

\bibitem{Beisert:2004ry}
N.~Beisert, {\it {The Dilatation operator of $\mathcal{N}=4$ super Yang-Mills
  theory and integrability}},  {\em Phys. Rept.} {\bf 405} (2004) 1--202,
  [\href{http://arxiv.org/abs/hep-th/0407277}{{\tt hep-th/0407277}}].

\bibitem{Minahan:2010js}
J.~A. Minahan, {\it {Review of AdS/CFT Integrability, Chapter I.1: Spin Chains
  in $\mathcal{N}=4$ Super Yang-Mills}},  {\em Lett.Math.Phys.} {\bf 99} (2012)
  33--58, [\href{http://arxiv.org/abs/1012.3983}{{\tt arXiv:1012.3983}}].

\bibitem{Adamo:2011dq}
T.~Adamo, M.~Bullimore, L.~Mason, and D.~Skinner, {\it {A Proof of the
  Supersymmetric Correlation Function / Wilson Loop Correspondence}},  {\em
  JHEP} {\bf 1108} (2011) 076, [\href{http://arxiv.org/abs/1103.4119}{{\tt
  arXiv:1103.4119}}].

\bibitem{Paper3}
L.~Koster, V.~Mitev, M.~Staudacher, and M.~Wilhelm {\em To appear}.

\bibitem{Chicherin:2016fac}
D.~Chicherin and E.~Sokatchev, {\it {$\mathcal{N}=4$ super-Yang-Mills in LHC
  superspace. Part I: Classical and quantum theory}},
  \href{http://arxiv.org/abs/1601.06803}{{\tt arXiv:1601.06803}}.

\bibitem{Chicherin:2016fbj}
D.~Chicherin and E.~Sokatchev, {\it {$\mathcal{N}=4$ super-Yang-Mills in LHC
  superspace. Part II: Non-chiral correlation functions of the stress-tensor
  multiplet}},  \href{http://arxiv.org/abs/1601.06804}{{\tt arXiv:1601.06804}}.

\bibitem{Chicherin:2016soh}
D.~Chicherin and E.~Sokatchev, {\it {Demystifying the twistor construction of
  composite operators in $\mathcal{N}=4$ super-Yang-Mills theory}},
  \href{http://arxiv.org/abs/1603.08478}{{\tt arXiv:1603.08478}}.

\bibitem{Derkachov:2013bda}
S.~Derkachov, G.~Korchemsky, and A.~Manashov, {\it {Dual conformal symmetry on
  the light-cone}},  {\em Nucl.Phys.} {\bf B886} (2014) 1102--1127,
  [\href{http://arxiv.org/abs/1306.5951}{{\tt arXiv:1306.5951}}].

\bibitem{Adamo:2013cra}
T.~Adamo, {\it {Twistor actions for gauge theory and gravity}},
  \href{http://arxiv.org/abs/1308.2820}{{\tt arXiv:1308.2820}}.

\bibitem{Hodges:2009hk}
A.~Hodges, {\it {Eliminating spurious poles from gauge-theoretic amplitudes}},
  {\em JHEP} {\bf 1305} (2013) 135, [\href{http://arxiv.org/abs/0905.1473}{{\tt
  arXiv:0905.1473}}].

\end{thebibliography}\endgroup

\end{document}